\documentclass[preprint,prc,aps,superscriptaddress, nofootinbib,showpacs]{revtex4}

\usepackage[dvips]{graphicx}
\usepackage{amssymb}

\def\beq {\begin{equation}}
\def\eeq {\end{equation}}
\def\bea {\begin{eqnarray}}
\def\eea {\end{eqnarray}}
\def\ni {\noindent}
\def\nn {\nonumber}

\def\lp {\left( }
\def\rp {\right) }
\def\lb {\left[ }
\def\rb {\right] }
\def\lc {\left\{ }
\def\rc {\right\} }
\def\ra {\;\rangle }
\def\la {\langle\; }

\def\rar {\rightarrow}
\def\lrar {\leftrightarrow}

\def\qb {\bar{q}}

\def\kb {\bar{\k}}

\def\Ob {\bar{\Omega}}
\def\Rb {\bar{R}}
\def\Tb {\bar{T}}

\def\ct {\tilde{c}}

\def\dr {\partial}
\def\di {\dr_\m}
\def\ds {\dr_\m}

\def\rth {\sqrt{3}}
\def\sp {\!+\!}
\def\sm {\!-\!}
\def\cd {\!\cdot\!}

\def\so {{_{1/2}}}
\def\st {{_{3/2}}}

\def\cA {{\cal{A}}}
\def\cK {{\cal{K}}}
\def\cL {{\cal{L}}}
\def\cM {{\cal{M}}}
\def\cO {{\cal{O}}}

\def\cW {{\cal{W}}}

\def\a {\alpha }
\def\d {\delta}
\def\D {\Delta}

\def\k {\kappa}
\def\f {\phi}
\def\g {\gamma}
\def\l {\lambda }
\def\m {\mu}
\def\n {\nu}
\def\o {\omega}
\def\p {\pi}
\def\s {\sigma}

\begin{document}

\title
{Decay $D^+ \rar K^-\p^+ \, \p^+$: chiral symmetry and scalar resonances}

\author{D. R. Boito}
\email[]{boito@ifae.es}

\affiliation{Grup de F\'isica Te\'orica and IFAE, Universitat Aut\`onoma 
de Barcelona, E-08193 Bellaterra (Barcelona), Spain.}

\author{P. C. Magalh\~aes}
\email[]{patricia@if.usp.br}
\author{M. R. Robilotta}
\email[]{robilotta@if.usp.br}
\author{G. R. S. Zarnauskas}
\email[]{gabrielz@if.usp.br}

\affiliation{Instituto de F\'{\i}sica, Universidade de S\~{a}o Paulo,\\
C.P. 66318, 05315-970, S\~{a}o Paulo, SP, Brazil.}

\date{\today}

\begin{abstract}
The low-energy $S$-wave component of the decay $D^+ \! \rar K^- \p^+ \p^+ $ 
is studied by means of a chiral $SU(3)\times SU(3)$ effective theory. 
As far as the primary vertex is concerned, we allow for the possibility
of either direct production of three pseudoscalar mesons or a meson
and a scalar resonance. 
Special attention is paid to final state interactions associated with
elastic meson-meson scattering. 
The corresponding two-body amplitude is unitarized by ressumming 
$s$-channel diagrams and can be expressed in terms of the 
usal phase shifts $\d$.
This procedure preserves the chiral properties of the amplitude 
at low-energies. 
Final state interactions also involve another phase $\o$, 
which describes intermediate two-meson propagation and is theoretically 
unambiguous.
This phase is absent in the $K$-matrix approximation.
Partial contributions to the decay amplitude involve a real term,
another one with phase $\d$ and several others with phases $\d+\o$.
Our main result is a simple and almost model independent 
chiral generalization of the usual Breit-Wigner expression, 
suited to be used in analyses
of production data involving scalar resonances.

\end{abstract}


\maketitle

\section{introduction}

Decays of heavy mesons have been recognized recently as important sources 
of information about scalar resonances, since the E791 experiment
has produced solid evidence for a broad scalar-isoscalar
state in $D^+\to \pi^+\pi^-\pi^+ $ \cite{E791}, know as the $\sigma$.
In the case of the reaction we are concerned with, namely 
$D^+ \! \rar K^- \p^+ \p^+ $, the E791 group included in their fit 
another scalar state, known as the $\kappa$ (or $K^*(800)$),
and concluded that this resonance was the dominant source
of $S$-wave $\pi^+ K^-$ pairs\cite{E791K}. 
This represented a turn point in our understanding of this reaction, 
which was hitherto thought to be dominated by the
non-resonant background. The existence of the $\kappa$  was then confirmed
in a different approach by FOCUS \cite{FOCUS} and by the new 
high statistics results from CLEO \cite{CLEOc}.
Empirical information about the parameters of these resonances, especially 
the position of their poles in the complex energy plane, can only be obtained
after analyses of large amounts of data organized into Dalitz plots.

Low-energy mesons correspond to gentle deviations from the QCD 
vacuum and tend to be highly collective states, as indeed happens with pions
or kaons, which are large objects.
In the case of low-energy scalar resonances, one is entitled to 
expect that 
the influence of the vacuum will be a truly overwhelming one.
This is probably one of the reasons why they are so broad
and prove to be so elusive.

Dalitz plots of $D$-meson decays into three pseudoscalars usually contain
several $S$- and $P$-wave resonances\cite{E791,Meadows}, which couple 
by means of final state interactions (FSIs).
This makes the disentangling of individual resonance properties both very 
involved and strongly dependent on particular forms of trial functions 
adopted.
Theoretical ans\"atze employed in data interpretation usually rely on 
Breit-Wigner expressions, which proved to work well in many instances.
A popular ansatz for the trial function has the form
\beq
\cA= a_0 \; e^{i\phi_0}\; \cA_0 
+ \sum_{\mathrm{S-wave}} \; a_n^S \; e^{i \phi_n^S}\; \cA_n^S 
+ \sum_{\mathrm{P-wave}} \; a_n^P \; e^{i \phi_n^P}\; \cA_n^P 
+ \cdots \;, 
\label{1.1}
\eeq

\ni
where the first term represents a non-resonant background and the amplitudes
$\cA_n^\ell$ are Breit-Wigner expressions for each resonance present in the 
final state.
The masses and widths of well established states are used as input,
whereas those of low-lying resonances, as well as the free parameters 
$a_i^\ell$ and $\phi_i^\ell$, are fitted to data.
As a consequence, these adjusted parameters acquire the status of empirical
quantities.

In the case of $\p \p$ scattering, the fact is well established
that the lowest pole of the amplitude is located roughly at 
$\sqrt{s} \approx (0.47 - i\, 0.29)$ GeV,
whereas the $S$-wave phase shift reaches $\p/2$ 
around $\sqrt{s_{_{\p/2}}} \approx 0.92$ GeV.
These findings are clearly at odds with the traditional use of 
Breit-Wigner expressions. 
An explanation for these seemingly paradoxical results was provided by 
Colangelo, Gasser and Leutwyler\cite{CGL}, who have shown that chiral 
symmetry requires a compromise between the polynomial nature of the 
amplitude at very low-energies and the vanishing of its real part at 
$s_{_{\p/2}}$, which is responsible for a large shift in the pole position.
This kind of feature is inherent to the isoscalar channel.
The resonance, which is a non-leading chiral effect, must always coexist 
with an important polynomial in $s$.
In the framework of chiral symmetry, the first two terms in 
eq.(\ref{1.1}), namely 
$[a_0 \; e^{i\phi_0}\; \cA_0 + a_1^S \; e^{i \phi_1^S}\; \cA_1^S]$,
are not suited for describing low-energy interactions.
The use of this kind of trial function is problematic and may give rise to  
results which are not reliable.

The extraction of information from experiments involving scalar resonances 
must be performed in the best theoretical framework 
possible, as the quality of results for masses and coupling constants 
depend on the ans\"atze employed.
In this work we propose an alternative form for the low-energy sector 
of the trial function, which can be used as a tool in 
analyses of the decay $D^+ \! \rar K^- \p^+ \p^+$.
Our motivation for choosing this particular reaction is two-fold.
The first one is that scalar resonances in the final state occur just 
in the $\p^+ K^-$ subsystem, the number of possible couplings is relatively 
small, and the problem is simplified.
The second is the availability of recent data analyses on this 
process, which could allow the testing of our results.
Nevertheless, the general lines adopted here can be extended
to other systems in a straightforward manner. 
Our theoretical model is based on standard $SU(3)\times SU(3)$ effective
chiral lagrangians incorporating scalar resonances\cite{GL,EGPR}.

We are also concerned with the presence of phases in theoretical ans\"atze,
which are sometimes employed with no visible physical meaning.
In the realm of two-body systems, the dynamical origin of 
phases is well understood.
In particular, it is known that two-body rescattering gives rise 
to the elastic phase shift $\d$.
When resonances are present, this phase is shared with the so called 
production amplitude, as dictated by Watson's theorem \cite{Watson}.
We show, in the sequence, that another 
phase is also present
in the $D^+$ decay considered here, associated with two-meson intermediate 
states.

The theoretical description of a heavy-meson decay into three 
pseudoscalars is necessarily complex.
Nevertheless, we have made an effort to produce final expressions 
which incorporate a compromise between
reliability and simplicity, so that they could be employed directly
in data analyses. 
Our paper is organized as follows.
In section II, we review the basic lagrangians, which are used in the 
description of both the primary weak vertex and final state interactions.
The former is discussed in section III, whereas the latter are discussed 
in sections IV, V and VI. 
These results are assembled in section VII, where our main results 
are presented and analyzed. 
In section VIII, we display individual predictions from the various 
components of the decay amplitude in Dalitz plots, so as to produce
a feeling of their dynamical content. 
Finally a summary and comprehensive conclusions are presented in section IX.
Details concerning kinematical variables, the form of the two-body propagator
and background interactions are left to appendices.

\section{dynamics}

The reaction $D^+ \! \rar K^-  \p^+ \p^+$ involves both weak and strong
processes.
The former are associated with the isospin conserving quark transition 
$c \rar s \, W^+$, whereas the latter occur in final state 
interactions involving both pseudoscalar mesons and their resonances.
In order to keep approximations under control, we remain, 
as much as possible, within the single theoretical framework provided by a
chiral effective field theory.
This choice is also motivated by the fact that we are concerned mostly
with the low-energy sector of the trial function.

\vspace{2mm}
\ni
{\bf $\bullet$ strong interactions:}

Meson-meson interactions are described by the chiral $SU(3)\times SU(3)$
lagrangian at $\cO(q^2)$ given by Gasser and Leutwyler\cite{GL},
whereas couplings of scalar resonances to pseudoscalar mesons are 
taken from the work of Ecker, Gasser, Pich and De Rafael\cite{EGPR}. 
Keeping only relevant terms, one has
\bea
\cL^{(2)} &\!=\!& \frac{F^2}{4}\; 
\la \nabla_\m \,U^\dag \; \nabla^\m \,U + \chi^\dag\,U + \chi\,U^\dag \ra
\nn\\
&\!+\!& c_d \;  \la S \; u_\m \; u^\m  \ra 
+ \; c_m \;  \la  S\; \chi_+ \ra 
\nn\\
&\!+\!& \ct_d \; S_1 \; \la  u_\m \; u^\m  \ra 
+ \; \ct_m \; S_1\; \la \chi_+ \ra ,
\label{2.1}
\eea

\ni
where $\la\cdots\ra$ indicates the trace, $F$, $c_d$, $c_m$, $\ct_d$ 
and $\ct_m$ are constants, $S$ and $S_1$ represent scalar resonances and 
$U$ is the pseudoscalar field.
Using the definition $u = U^{1/2}$, one has
\bea
\nabla_\m \,U = \dr_\m \,U \;,
&\;\;\;\;\;&
u_\m  = i\, u^\dagger \; \nabla_\m U \; u^\dagger \;,
\nn\\
\chi = 2\,B\, \s \;,
&\;\;\;\;\;&
\chi_+ = u^\dagger \; \chi \; u^\dagger + u \; \chi^\dagger \; u \; .
\label{2.2}
\eea

\ni
The field $\s$ incorporates the quark masses as external scalar sources and,
in the isospin limit, is written in terms of the usual Gell-Mann matrices 
as $\s = \sigma_0 \; I + \sigma_8 \; \l_8$, with
$\s_0 = (2\, \hat{m} \sp m_s)/3$, $\s_8= (\hat{m} \sm m_s)/\rth $
and $m_u=m_d\equiv \hat{m}$.
We neglect $\eta$-$\eta'$ mixing and use $B\,\s_0=(2\,M_K^2 \sp M_\p^2)/6$
and $B\,\s_8=-(M_K^2 \sm M_\p^2)/\sqrt{3}$.
The meson field is written as
$\; U \equiv e^{i \Phi/F}, \;\; \Phi \equiv \l_i \; \phi_i \,$,
and the leading order strong lagrangian becomes
\bea
\cL^{(2)}&=& -\, \frac{1}{6 F^2}\,f_{ijs}\,f_{kls}\,
\f_i\,\di\f_j\, \f_k\, \ds\f_l  
+ \frac{B}{24 F^2} \, \lb \s_0 \, 
\lp \frac{4}{3}\,\d_{ij}\,\d_{kl} \sp  2\, d_{ijs}\,d_{kls} \rp \right.
\nn\\[3mm]
&+& \left. \s_8 \; 
\lp \frac{4}{3}\,\d_{ij}\,d_{kl8} \sp \frac{4}{3}\, d_{ij8}\,\d_{kl} 
\sp 2\, d_{ijm}\, d_{kln}\, d_{8mn} \rp  \rb \, \f_i\,\f_j\,\f_k\,\f_l 
\nn\\[3mm]
&+& \frac{2\, \ct_d}{F^2}\; S_0 \; \di \f_i \; \ds \f_i 
- \frac{4 \, \ct_m}{F^2} \,B\;
\lb \s_0 \, \d_{ij} + \s_8 \, d_{8ij} \rb \, S_0 \, \f_i\, \f_j
\nn\\[3mm]
&+& \frac{2\, c_d}{F^2}\;d_{bij}\; S_b \; \di \f_i \; \ds \f_j 
- \frac{c_m}{F^2} \,B\;
\lb 3\,\s_0 \, d_{aij}
+ \s_8 \, \lp 2\, \d_{bi} \,\d_{j8} + 4 \, d_{bis}\,d_{sj8} \rp \rb 
\, S_b \, \f_i\, \f_j \;,
\label{2.3}
\eea

\ni
where $f_{ijk}$ and $d_{ijk}$ are the usual $SU(3)$ constants.
We use the conventions of ref.\cite{G} for meson fields and 
$\kb^0 \equiv (S_6 \sp i S_7)/\sqrt{2}$ for the scalar resonance.
In the sequence, the $\kb^0$ state is called $\k$ for simplicity and,
in isospin space, one has
\bea
&& | \p^+ K^- \ra= \sqrt{1/3} \; |3/2, 1/2 \ra 
+ \sqrt{2/3} \; |1/2, 1/2 \ra \;,
\nn\\
&& | \p^0 \bar{K}^0 \ra= \sqrt{2/3} \; |3/2, 1/2 \ra 
- \sqrt{1/3} \; |1/2, 1/2 \ra\;.
\label{2.4}
\eea

\vspace{2mm}
\ni
{\bf $\bullet$ weak interactions:}

\begin{figure}[h] 
\includegraphics[width=0.9\columnwidth,angle=0]{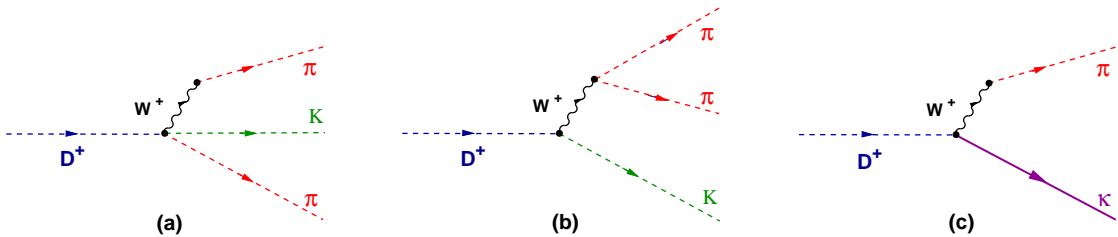}
\caption{(Color online) Weak amplitudes involving pseudoscalar mesons 
(dashed lines) and the scalar resonance (continuous line).}
\label{F1}
\end{figure}

Effective weak vertices contain a propagating $W$ and are based on the 
processes shown in fig.\ref{F1}, which involve the 
following combinations of reactions:\\
type (a): $\;\; (D^+ \rar \p \, K \, W^+)_a$ and $(W^+ \rar \p)_a \;$ ,\\
type (b): $\;\; (D^+ \rar K \, W^+)_v$ and $(W^+ \rar \p \, \p)_v \;$ ,\\
type (c): $\;\; (D^+ \rar \kb \, W^+)_a$ and $(W^+ \rar \p)_a \;$ ,\\
where the labels $v$ and $a$ refer to either vector or axial currents 
in $W$ couplings.
In the want of a comprehensive theory, these weak vertices can be 
derived by means of appropriate hadronic currents and {\em ad hoc} 
phenomenological coupling constants.
The latter can be estimated in semi-leptonic processes, by replacing
the top $\p^+$ in the figure by $(\ell^+ \; \n_\ell)$.
Although feasible, the piecemeal implementation of this program is 
cumbersome, owing to the large number of possible isospin couplings
and phase conventions.

Fortunately a rather economic alternative is available, based on the 
group $SU(4)$.
Before proceeding, we would like to make clear that we are quite aware 
that $SU(4)$ is {\em not} a good symmetry and therefore we are {\em not} 
advocating its use here.
On the other hand, the inclusion of $D$ mesons into the pseudoscalar 
multiplet does give rise, by means of the external source technique, 
to an effective lagrangian in which all currents and isospin couplings
are generated automatically. 
We obtain $W^+$ couplings by using a $SU(4)\times SU(4)$ mesonic 
lagrangian for the weak sector and, at the end, break any commitments 
with a {\em symmetry}, by allowing phenomenological coupling constants
into the Feynman rules.

In the standard model lagrangian, weak couplings are given by 
$\cL_w = \qb \, \g_\m \lb v_\m + \g_5\, a^\m \rb q $,
where the external sources are written in terms of $SU(4)$ matrices 
as  $v^\m = v_a^\m \; \l_a/2$ and 
$a^\m = a_a^\m  \; \l_a/2$.
The charged currents are
\beq
v_\m + a_\m = 0\;, 
\;\;\; \lrar \;\;\; 
v_\m - a_\m  = -\,\frac{g}{\sqrt{2}}\;
\lb W_\m^+ \; V_{SU(4)} + W_\m^- \; V_{SU(4)}^\dagger \rb  \;,
\label{2.5}
\eeq

\ni
where $g$ is the weak coupling constant, $W$ is the gauge field and 
$V_{SU(4)}$ is the $SU(4)$ sector of the 
Cabibbo-Kobayashi-Maskawa mixing matrix.
Its explicit form, in terms of the Cabibbo angle $\theta_C$, 
reads\footnote{We use the $SU(4)$ conventions of ref.\cite{GM}.} 
\bea
V_{SU(4)} &\!=&\! \frac{1}{2} \;
\lb \cos\theta_C \; \lp \l_1 + i\, \l_2 \rp 
+ \sin\theta_C \; \lp \l_4 + i\, \l_5 \rp \right.
\nn\\
&\!-&\!\left. \sin\theta_C \; \lp \l_{11} + i\, \l_{12} \rp 
+ \cos \theta_C \; \lp \l_{13} + i\, \l_{14} \rp \rb .
\label{2.6}
\eea

The weak effective lagrangian is obtained by replacing $\nabla_\m$ 
in eqs.(\ref{2.1}) and (\ref{2.2}) with
$\nabla_\m \,U = \dr_\m \, U - i \;(v_\m \sp a_\m) U 
+ i\;U (v_\m \sm a_\m) $.
The term that interests is then given by
\bea
\cL_w^{(2)}&\!=\!& 
v_a^\m \lb f_{ajk}\; \f_j \, \di \f_k \rb
- a_a^\m \lb
F \, \d_{al} \, \di\f_l 
+ \frac{2}{3\,F} \; f_{ajs}\, f_{kls} \, \f_j \, \f_k \, \di\f_l 
+ \frac{4\,c_d}{F} \; d_{abl}\; S_b \, \di\f_l \rb,
\label{2.7}
\eea

\ni
where the factors within square brackets are respectively vector
and axial hadronic currents.

\section{weak vertices}

In this section we display the tree level amplitudes given in fig.\ref{F1}
and include final state interactions afterwards.
The weak vertices in which the $D^+$ participates 
involve the transformation of a $c$-quark into a $s$-quark 
and are Cabibbo allowed.
As final state interactions allow the transition $\p^0 \bar{K}^0 \rar \p^+ K^-$,
one also needs to consider vertices involving neutral mesons.
The production of a single pion in $W_\m^+ \rar \p^+(q')$
is represented by 
\beq
T_w^{\p} = i\; (F\,g/2) \; \cos \theta_C \; q'_\m \;.
\label{3.1}
\eeq

Decays of the type (a) are based on the processes 
$D^+(P) \rar W_\m^+ \; K^-(k) \; \p^+(q)$ 
and
$D^+(P) \rar W_\m^+ \; K^0(k) \; \p^0(q)$, 
given respectively by the amplitues 
$\sqrt{2/3} \; T_w^{D\p K}$ and $ - \sqrt{1/3} \; T_w^{D\p K}$, 
with
\beq
T_w^{D\p K} = -i \; (g/2\sqrt{6}\;F)\; \cos \theta_C \; 
(P - k)_\m \;.
\label{3.2}
\eeq

In the case of vector couplings, one needs the vertices 
$D^+(P) \rar W_\m^+ \; K^0(k)$ and $W_\m^+ \rar \p^+(q') \; \p^0(q)$,
which read
\bea
&& T_w^{DK} = - (g/2 \sqrt{2}) \; \cos \theta_C \; (P + k)_\m \;,
\label{3.3}\\
&& T_w^{\p\p} =  (g/2) \; \cos \theta_C \; (q - q')_\m \;.
\label{3.4}
\eea

Finally, for the vertex $D^+(P) \rar W_\m^+ \; \bar{\k}$, 
involving the scalar resonance, one has 
\beq
T_w^{D \bar{\k}} = i (c_d\,g \sqrt{2}/F)\; 
\cos\theta_C \; P_\m \;.
\label{3.5}
\eeq

Using $\D_W^{\m\n} = i \;( g^{\m \n}/M_W^2)$ for the $W$ propagator 
and the definition $G_F \equiv \sqrt{2}\,g^2/8M_W^2=
1.166\times 10^{-5}\,$GeV$^2$\cite{PDG}
, one finds the following
weak amplitudes\\
type (a):
\bea
&& D^+(P) \rar K^-(k) \; \p^+(q)\; \p^+(q'):
\;\;\;\sqrt{2/3}\;\; \cW_a \;,
\label{3.6}\\
&& D^+(P) \rar \bar{K}^0(k) \; \p^0(q)\; \p^+(q'):\;\;\;
- \sqrt{1/3}\;\; \cW_a\;,
\label{3.7}\\
&& \cW_a = [\d_a]\,
(G_F/\sqrt{3})\;\cos^2\theta_C \; (P - k) \cdot q' \;,
\eea
type (b):
\beq
D^+(P) \rar \bar{K}^0(k) \; \p^0(q)\; \p^+(q'):\;\;\;
\cW_b = - [\d_b] \, G_F \; \cos^2\theta_C \; (P + k) \cdot (q - q')\;,
\label{3.8}
\eeq
type (c):
\beq
D^+(P) \rar \bar{\k}(q_s) \; \p^+(q'):\;\;\;
\cW_c = -\,[\d_c]\, 4\; G_F\; c_d \; \cos^2\theta_C \; P \cdot q' \;,
\label{3.9}
\eeq

\ni
where  {\em ad hoc} factors $[\d_i]$ were introduced so to freeing  
results from any constraints imposed by $SU(4)$ symmetry.

\section{FSI: kernel}
\label{kernel}

Final state interactions are essential to structures observed 
in Dalitz plots, since they promote couplings among various channels and, 
in particular, give rise to widths of resonances.
The final state considered here contains three mesons
and a complete treatment of the problem is not possible.
One is forced to employ approximations and we adopt the 
quasi-two-body approach, in which one of the final mesons acts as 
a mere spectator.
As emerging pions have isospin 2 and no resonance is known in this 
channel, their interactions can be safely 
neglected and strong interactions are restricted to the $\p K$ subsystem.

\begin{figure}[h] 
\includegraphics[width=0.8\columnwidth,angle=0]{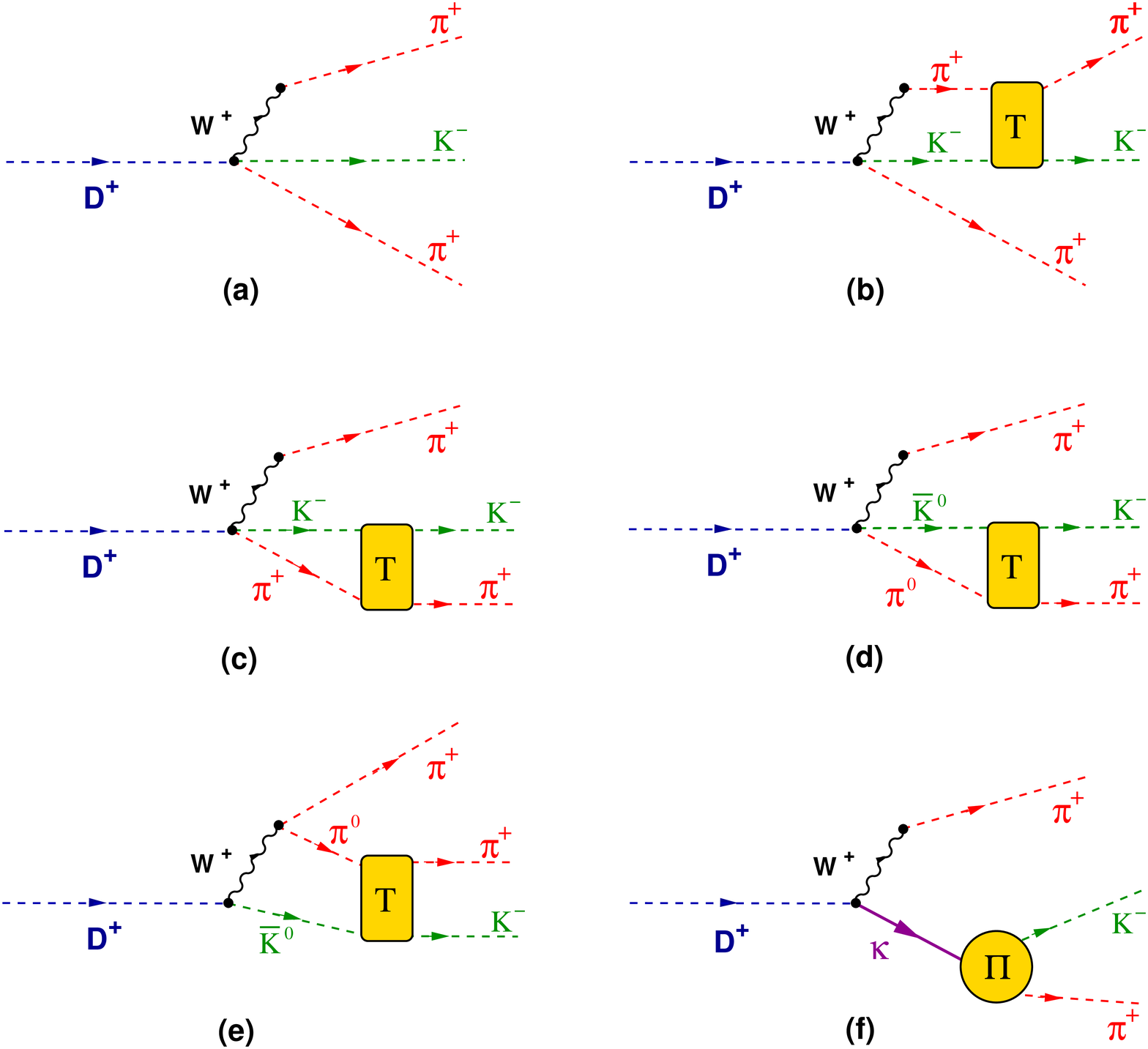}
\caption{(Color online) Diagrams contributing to the decay
$D^+ \! \rar K^- \p^+ \p^+$; (a) corresponds to a direct process,
(b-e) involve the $\p K$ scattering amplitude $T$,
and (f) depends on the production amplitude $\Pi$.}
\label{F2}
\end{figure}

\begin{figure}[h] 
\includegraphics[width=1.0\columnwidth,angle=0]{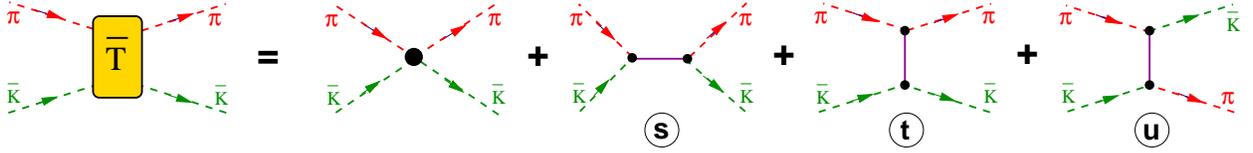}
\vspace{-12mm}
\caption{(Color online) Tree-level $\pi K$ amplitude;
dashed and full lines represent respectively pseudoscalar mesons 
and  scalar resonances.}
\label{F3}
\end{figure}

These assumptions lead to the model given by fig.\ref{F2}, 
in which all strong processes are incorporated into the amplitudes 
$T$ and $\Pi$, representing respectively elastic scattering and production.
As the isospin of the $\p K$ system can be either $1/2$ or $3/2$ and only
the former couples with the $\k$, one needs to consider three amplitudes,
namely $T_\so$, $T_\st$ and $\Pi_\so$.
The basic building blocks of both $T_I$ and $\Pi_I$ are kernels $\cK_I$,
which describe elastic $S$-wave $\p K$ scattering at tree level.
These kernels are obtained by projecting out $S$-wave components from  
tree-level amplitudes $\Tb_I$, given by the diagrams of fig.\ref{F3}.
As far as chiral symmetry is concerned, the leading term is given by
the $\cO(q^2)$ contact term whereas diagrams involving resonances are 
$\cO(q^4)$ corrections.

We consider the process $\p(Q)\, K(K) \rar \p(q)\, K(k)$ and
the tree amplitudes are explicitly written in terms of the Mandelstam
variables as 
\bea
\Tb_\so \!\! &=& \!\!  
\frac{1}{4 \,F^2} \,\lb 4\,s + 3\, t - 4\,(M_\p^2 + M_K^2)\rb
\nn\\[2mm]
&-& \!\! \frac{3}{4}\; \frac{1}{s \sm m_\k^2}\; \frac{4}{F^4}\;
[c_d\;(s \sm M_\p^2 \sm M_K^2) + c_m \;(5\,M_\p^2 \sp 4\,M_K^2)/6 ]^2
\nn\\[2mm]
&+& \!\! \Tb_t^0 -\Tb_t^8/6 - \Tb_u^\k/4 \;,
\label{4.1}\\[2mm]
\Tb_\st \!\! &=& \!\!  
-\, \frac{1}{2 \,F^2} \,\lb s - (M_\p^2 + M_K^2)\rb 
+ \Tb_t^0 - \Tb_t^8/6 + \Tb_u^\k/2 \;,
\label{4.2}\\[4mm]
\eea
\ni
where
\bea
\Tb_t^0 \!\! &=& \!\!  
- \, \frac{1}{t \sm m_0^2}\; \frac{4}{F^4}
\lc [\ct_d\,(t \sm 2\,M_\p^2) + 2\, \ct_m \, M_\p^2] \;
[\ct_d\, (t \sm 2\, M_K^2)+ 2\,\ct_m \, M_K^2] \rc \;,
\label{4.3}\\
\Tb_t^8 \!\! &=& \!\!
-\;\frac{1}{t \sm m_8^2}\; \frac{4}{F^4}
\lc [ c_d\;(t \sm 2 M_\p^2) - c_m \;(2 M_K^2 \sm 11 M_\p^2)/6] \right.
\nn\\
&\times & \!\! \left.
[ c_d\;(t \sm 2 M_K^2) + c_m \;(10 M_K^2 \sm M_\p^2)/6] \rc \;,
\label{4.3bis}\\
\Tb_u^\k \!\! &=& \!\! 
- \, \frac{1}{u \sm m_\k^2}\; \frac{4}{F^4}\;
[c_d\;(u \sm M_\p^2 \sm M_K^2) + c_m \;(4\,M_K^2 \sp 5\, M_\p^2)/6 ]^2
\;,
\label{4.4}
\eea

\ni
The projection into $S$-waves is performed using results
from appendix \ref{kinematics} and one finds
\bea
\cK_\so \!\! &=& \!\!  
\frac{1}{4 \,F^2} \,\lb \lp 4 -3\,\rho^2/2 \rp s
- 4\, \lp M_\p^2 + M_K^2\rp\rb
\nn\\[2mm]
&-& \!\! \frac{1}{s \sm m_\k^2}\; \frac{3}{F^4}\;
[c_d\, \lp m_\k^2 \sm M_\p^2 \sm M_K^2 \rp 
+ c_m \, \lp 5\,M_\p^2 \sp 4\,M_K^2 \rp /6 ]^2
+ B_\so\;,
\label{4.5}\\[2mm]
\cK_\st \!\! &=& \!\!  
-\, \frac{1}{2 \,F^2} \,\lb s - (M_\p^2 + M_K^2)\rb + B_\st\;,
\label{4.6}
\eea

\begin{figure}[h]
\hspace*{-5mm} 
\includegraphics[width=0.52\columnwidth,angle=0]{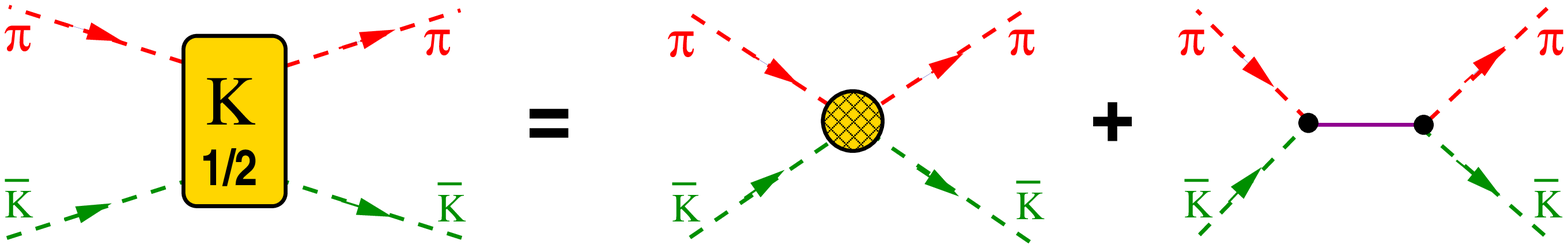}
\hspace*{15mm}
\includegraphics[width=0.35\columnwidth,angle=0]{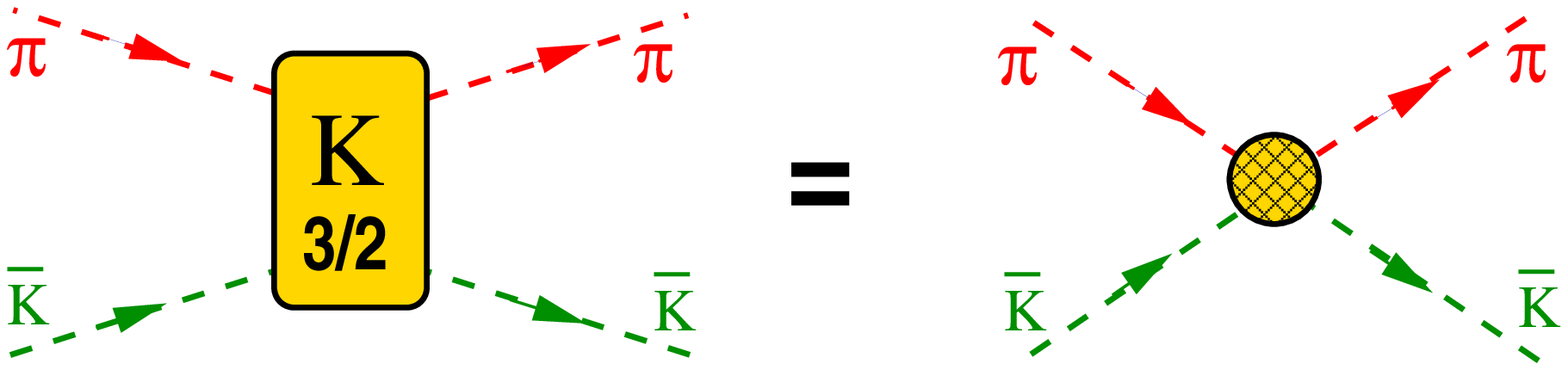}\\
\caption{(Color online) Effective structures of the kernels $\cK_I$;
the cross hatched bubbles represent effective contact interactions,
which include both the leading $\cO(q^2)$ contribution
and background terms.}
\label{F4}
\end{figure}

\ni with 
$\rho=\sqrt{1 - 2\, (M_K^2 \sp M_\p^2)/s+ (M_K^2 \sm M_\p^2)^2/s^2}$.
The functions $B_I$ are smooth backgrounds given explicitly in 
appendix \ref{background}.
As discussed there, all $t$ and $u$ channel are small and can be either
treated as four-point contact interactions or neglected.
This gives rise to the effective structures shown in fig.\ref{F4}, where 
contact terms in the kernels include both leading chiral contributions and 
non-resonant backgrounds.

\vspace{3mm}

In the isospin $1/2$ channel, it is convenient to emphasize the role 
of  the resonance by factorizing the $s$-channel  denominator and 
writing\cite{AR06}
\bea
\cK_\so \!\!&=&\!\! -\, \frac{\g^2}{s \sm m_\k^2} \;,
\nn\\[2mm]
&\g^2 \!\!&= (3/ F^4) \,
[c_d\; \lp m_\k^2 \sm M_\p^2 \sm M_K^2 \rp 
+ c_m \; \lp 5\,M_\p^2 \sp 4\,M_K^2 \rp /6 ]^2
\label{4.7}\\
&-& \!\!  \lc (1/ 4 F^2)\,
\lb \lp 4 -3\,\rho^2/2 \rp s - 4\, \lp M_\p^2 + M_K^2\rp \rb 
+ B_\so \rc \,(s \sm m_\k^2) \;.
\nn
\eea

\ni
Of course, in spite of differences in form, eqs.(\ref{4.5}) and (\ref{4.7}) 
have exactly the same content. 

As discussed in the introduction, we are interested in mapping 
low-energy degrees of freedom of the amplitude $D^+\! \rar K^+ \p^+ \p^+$.
This means that masses and coupling constants must be kept free, 
so that their values can be extracted from experiment. 
On the other hand, in discussing qualitative features of our results, 
we need to fix somehow these free parameters.
In this case, we choose:
$m_\k=1.2$ GeV, $(c_d, c_m)= (3.2, 4.2)\times 10^{-2}$ GeV,
$(\ct_d, \ct_m)= (1.8, 2.4)\times 10^{-2}$ GeV\cite{EGPR}.
It is worth stressing that we are by no means recommending these values.

\begin{figure}[h]
\hspace*{-15mm}
\includegraphics[width=0.45\columnwidth,angle=0]{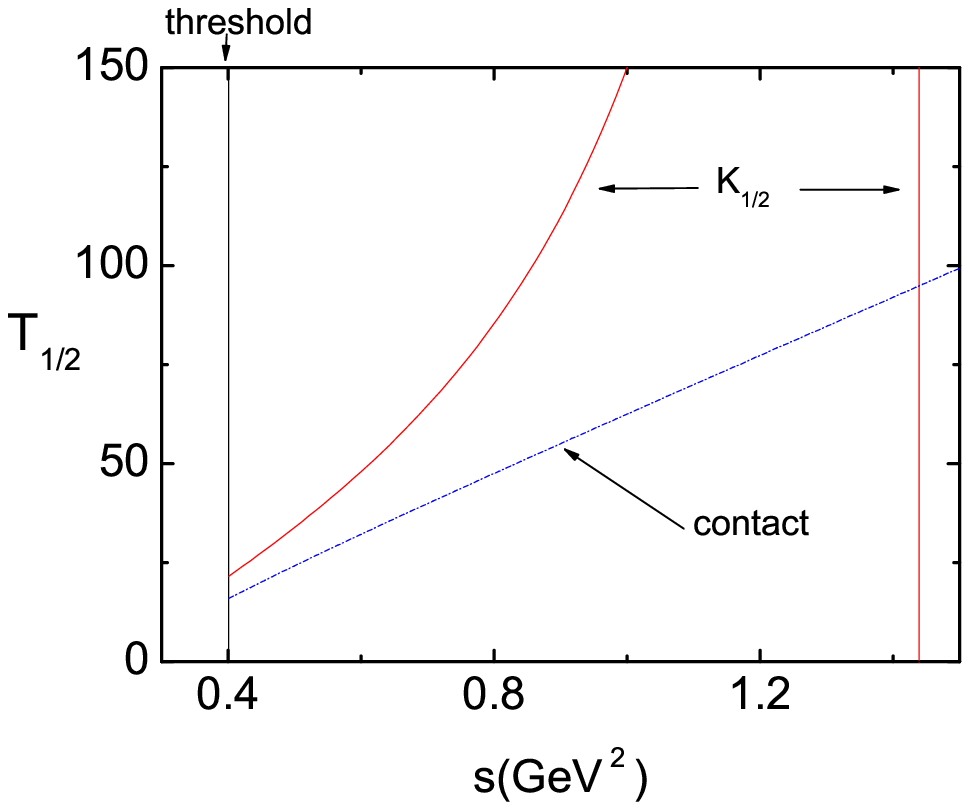}
\hspace*{10mm}
\includegraphics[width=0.45\columnwidth,angle=0]{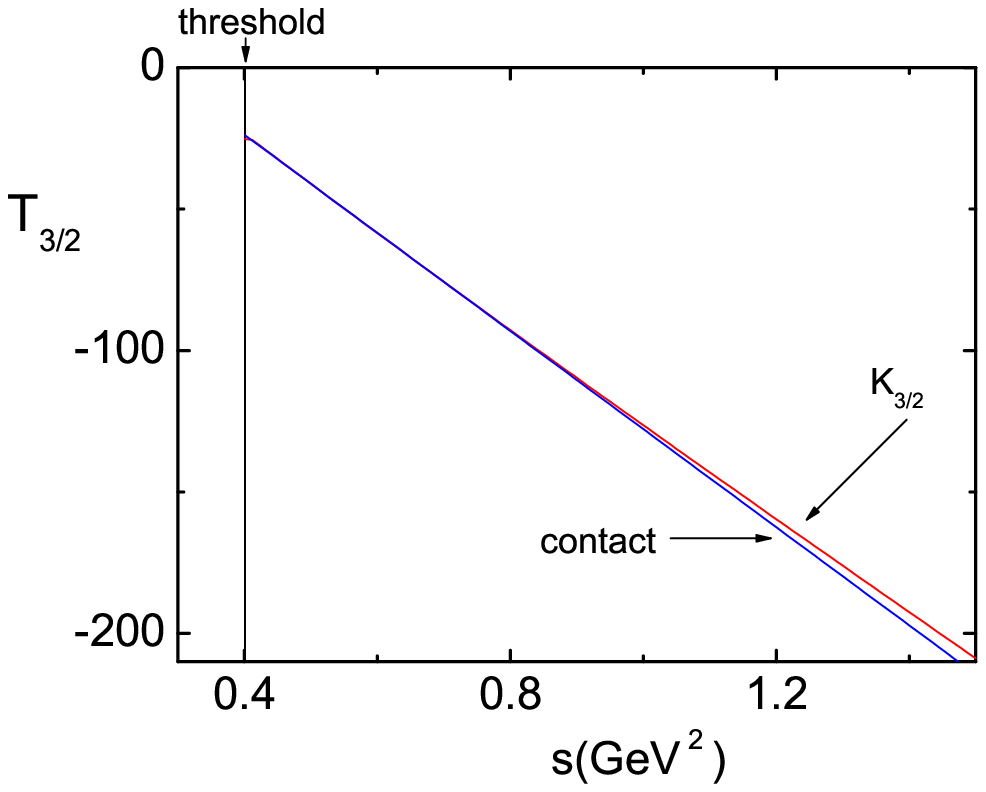}
\caption{(Color online) Kernels $\cK_I$ (continuous lines) and 
the leading $\cO(q^2)$ contact chiral contribution(dashed lines).}
\label{F5}
\end{figure}

In fig.\ref{F5} we display the full kernels $\cK_I$, together with their 
leading $\cO(q^2)$ chiral components, and it is possible to note 
an important isopin dependence of the results.  
The dynamical structure of $\cK_\so$ involves three different regimes.
At low energies, for $s$ between threshold and $\sim 0.6 \,$GeV$^2$,
it is determined by chiral constraints whereas,
as $s$ increases, it becomes dominated by the first resonance pole.
For larger values of $s$, effects associated with other resonances, not 
considered in this work, do show up.
Therefore, with the choice $m_\k=1.2\,$GeV, the upper limit of validity 
for our results is assumed to be $\sqrt{s} \sim 1.3\,$GeV.
The kernel $\cK_\st$ is repulsive and monotonic.

\section{FSI: scattering amplitude}
\label{scattering}

\begin{figure}[ht] 
\includegraphics[width=1.0\columnwidth,angle=0]{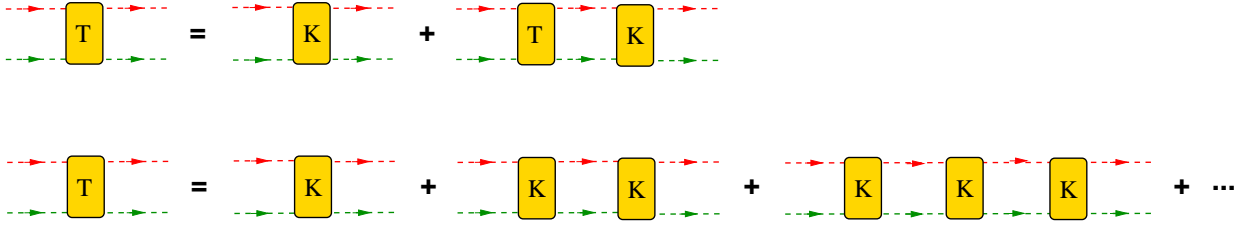}
\vspace{-8mm}
\caption{(Color online) Bethe-Salpeter equation for the elastic $\p K$
amplitude:  full equation (top) and perturbative solution (bottom).}
\label{F6}
\end{figure}

The elastic $\p K$ scattering amplitudes $T_I$ are derived by using
the two-body irreducible kernels $\cK_I$ into the Bethe-Salpeter equation, 
written schematically as
\beq
T_I = K_I + i \int \frac{d^4 \ell}{(2\p)^4}\; \cK_I(\ell) \; 
\D_{\p K}(\ell) \; T_I(\ell) \;,
\label{5.1}
\eeq

\ni
where $\D_{\p K}$ is the two-meson propagator.
The diagramatic representation of this equation, together with its 
perturbative solution, are shown in fig.\ref{F6}.
In the case of low-energy interactions, the treatment of the Bethe-Salpeter
equation
can be enormously simplified, as pointed out by Oller and Oset\cite{OO97}.
The fact that the kernels $\cK_I$ involve only effective contact interactions 
and $s$-channel resonances makes the two-meson propagator $\D_{\p K}$ 
to depend just on $s$ and eq.(\ref{5.1}) can be rewritten as
\bea
&& T_I = K_I - \cK_I \; \Omega \; T_I \;,
\label{5.2}\\[2mm]
&& \Omega = i\,\int \frac{d^4 \ell}{(2\p)^4}\; 
\frac{1}{[(q \sp k)/2 \sp \ell]^2 - M_\p^2}\;
\frac{1}{[(q \sp k)/2 \sm \ell]^2 - M_K^2} \;.
\label{5.3}
\eea

The function $\Omega$ diverges and has to be regularized.
As discussed in appendix \ref{propagator}, this brings into the
problem one free parameter for each isospin channel.
The corresponding regular functions are denoted by $\Ob_I$ and the 
solutions of eq.(\ref{5.2}) become
\beq
T_I = \frac{\cK_I}{1 + \Ob_I \, \cK_I} \;.
\label{5.4}
\eeq

\ni
Above threshold, the functions $\Ob_I$ are complex and written as 
$\Ob_I =  \Rb_I + i\, I \;$.
They involve a loop phase $\omega_I \equiv \tan^{-1} [I/\Rb_I]$
and their explicit analytic forms are given in appendix \ref{propagator}.

The amplitudes $T_I$ do respect unitarity and, below the first inelastic 
threshold, can always be written 
as\footnote{The amplitude $T$ is relativistic and we employ 
the conventions of Refs. \cite{Achasov} and \cite{AR06}.}
\bea
&& T_I = \frac{16 \p}{\rho} \,  \sin \d_I \, e^{i \d_I} \;,
\label{5.5}\\[2mm]
&& \;\;\;\;\; \tan\d_I = -\, \frac{I\;\cK_I}{1 + \Rb_I\;\cK_I}\;,
\label{5.6}
\eea

\ni
where the real phase shifts $\d_I$ incorporate
the dynamical content of the interaction.
These results allow the kernels to be expressed as 
\beq
\cK_I = \frac{ 16\pi}{\rho}\, \frac{\tan \delta_I} 
{(1+ \tan\delta_I/\tan\omega_I )}.
\label{5.7}
\eeq

\ni
The unitarization procedure employed in the derivation of $T_I$
generalizes that based on the on-shell iteration of the $K$-matrix 
\cite{Achasov,Black}, which amounts to neglecting the real part 
of $\Ob_I$ and to assuming
$\cK_I \simeq [16\pi/\rho] \, \tan \delta_I $.

The behaviors of the amplitudes $T_\so$ and $T_\st$ are very different, 
owing to the presence of a resonance in the former.
In this case, we use the kernel (\ref{4.7}) into eq.(\ref{5.4})
and finds
\beq
T_\so = \frac{16\p}{\rho}\; 
\frac{m_\k\,\Gamma_\k}{\cM_\k^2 - s -i\,m_\k \, \Gamma_\k} \;,
\label{5.8}
\eeq 

\ni
where the running mass and width are defined by
\bea
\cM_\k^2 \!&\equiv&\! m_\k^2 + \g^2 \,\Rb_\so \;,
\label{5.9}\\[2mm]
m_\k \,\Gamma_\k \!&\equiv&\! \g^2\,\rho/(16 \p) \;,
\label{5.10}
\eea

\ni 
and the free parameter in $\Rb_\so$ was chosen so that 
$\cM_\k^2(m_\k^2)=m_\k^2$.
This yields a unitary amplitude $T_\so$ which becomes purely 
imaginary at $s=m_\k^2$.
Therefore, we call $m_\k$ the {\em nominal} kappa mass.
Predictions for phase shifts can also be expressed as 
\beq
\tan \d_\so= \frac{m_\k \,\Gamma_\k}{\cM_\k^2 -s} 
\label{5.11}
\eeq

\ni
and the relationship between nominal and running masses is determined by 
\beq
\frac{m_\k^2 -s}{\cM_k^2 -s} = 1 + \tan \d_\so/\tan 
\omega_\so\;.
\label{5.12}
\eeq

\ni
$K$-matrix unitarization corresponds to making
$\cM_\k^2 \rar m_\k^2$ and deviations between both approaches
are quantified by the factor $(1 + \tan\d_\so / \tan\omega_\so)$.

\begin{figure}[h]
\hspace*{-40mm}
\includegraphics[width=0.6\columnwidth,angle=0]{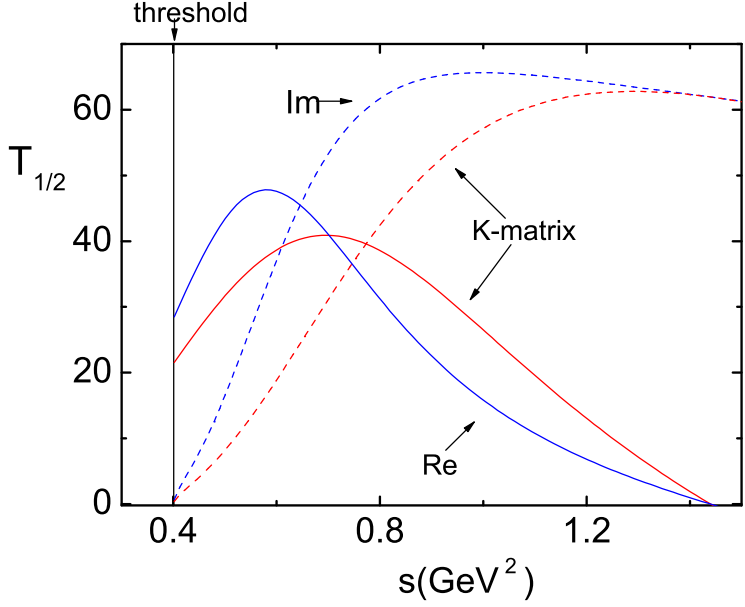}
\caption{(Color online) Real (full lines) and imaginary (dashed lines) 
components of the amplitude $T_\so$ obtained by means of eq.(\ref{5.8}) 
(blue) and in the $K$-matrix approximation (red).}
\label{F7}
\end{figure}

Real and imaginary parts of the amplitude $T_\so$, eq.(\ref{5.8}), 
together with the corresponding $K$-matrix approximation, 
are given in fig.\ref{F7}, for the choice $m_\k=1.2\;$GeV.  
In both cases, amplitudes become purely imaginary at $s=m_\k^2$
and peaks occur at lower energies.
As one discusses in section \ref{sum}, these shifts in the peaks are 
a direct consequence of chiral symmetry.
The figure shows that the $K$-matrix approximation is a crude one
and that the role played by the running mass is important.

\begin{figure}[h]
\hspace*{-15mm}
\includegraphics[width=0.45\columnwidth,angle=0]{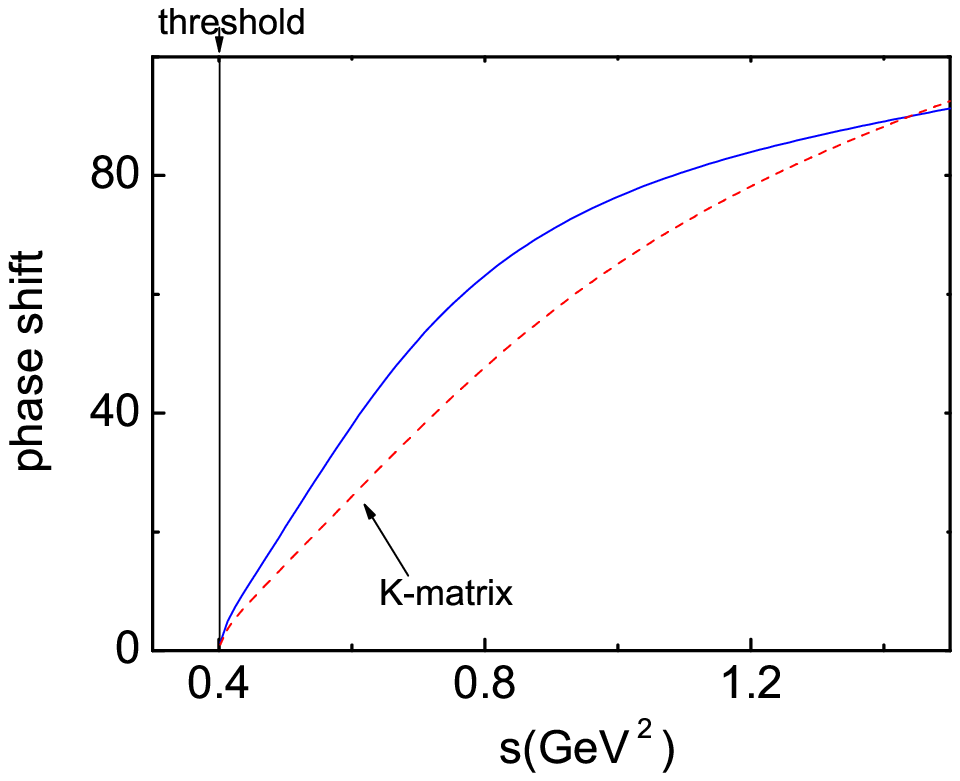}
\hspace*{5mm}
\includegraphics[width=0.45\columnwidth,angle=0]{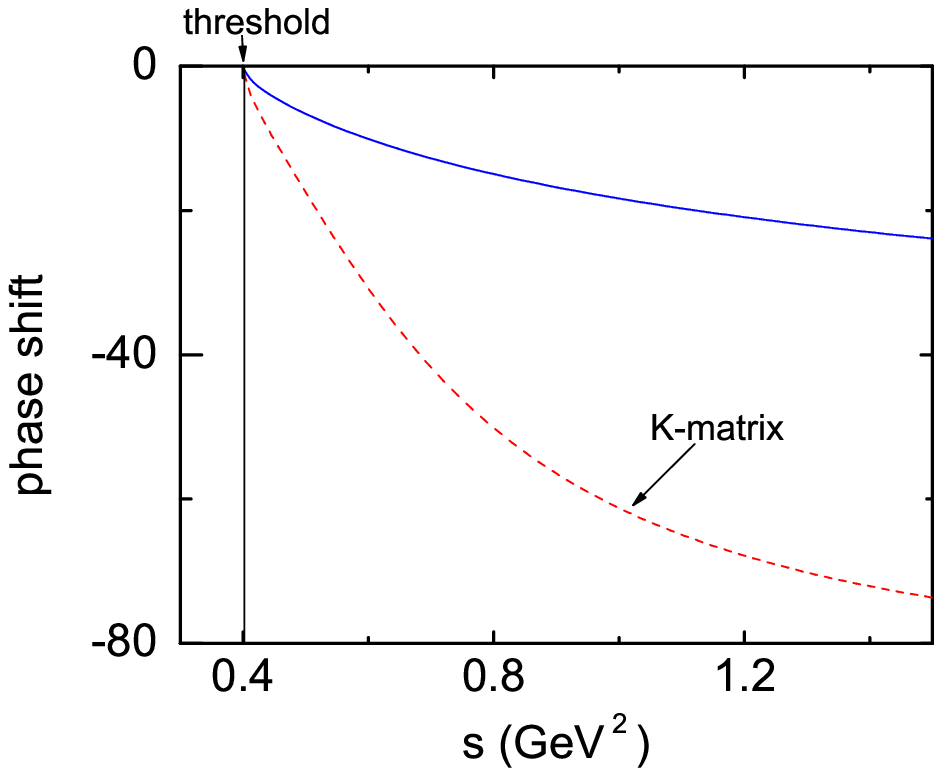}
\caption{(Color online) Predicted phase shifts for $T_\so$ and $T_\st$
amplitudes, together with $K$-matrix results.
}
\label{F8}
\end{figure}

\begin{figure}[h]
\hspace*{-6mm}
\includegraphics[width=0.45\columnwidth,angle=0]
{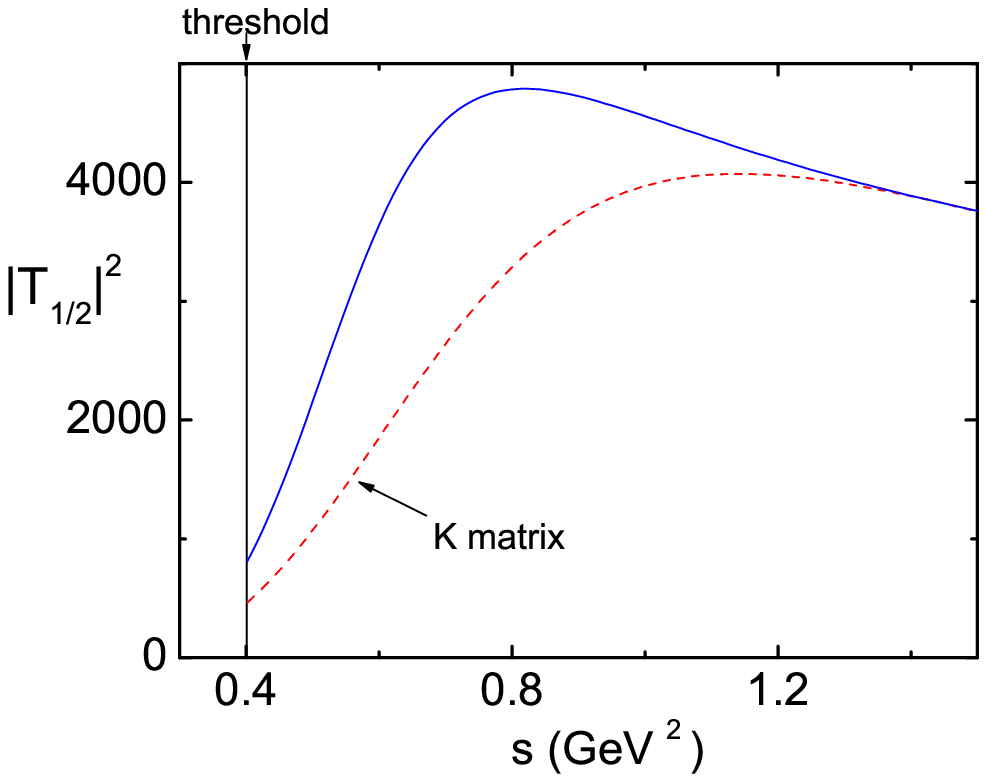}
\hspace*{10mm}
\includegraphics[width=0.45\columnwidth,angle=0]
{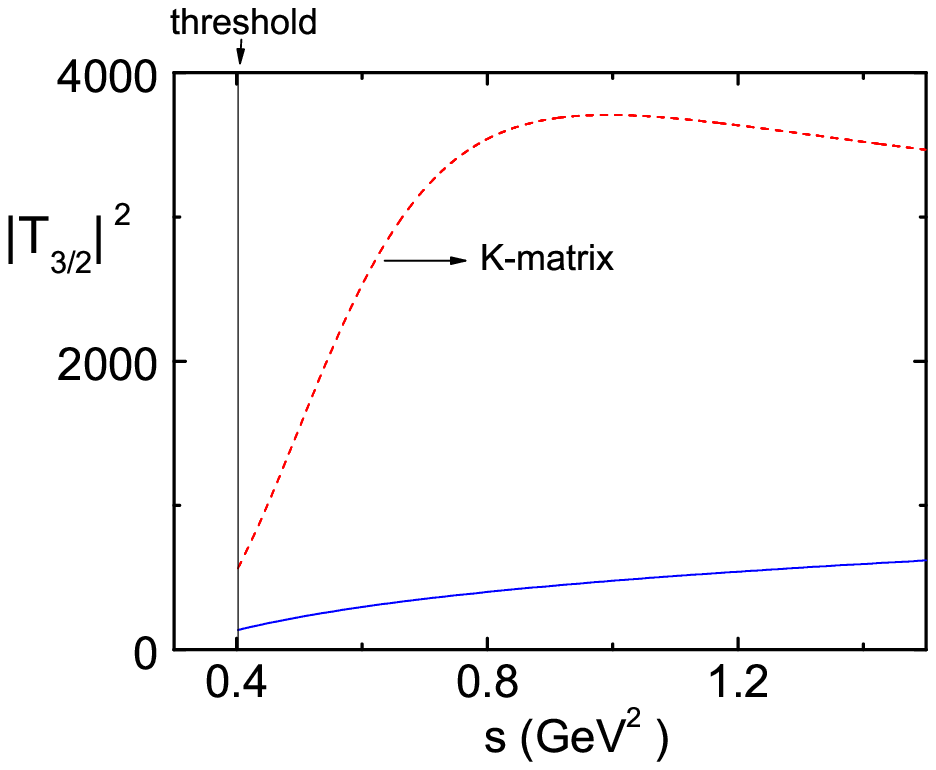}\\
\caption{(Color online) Dependence of $|T_\so|^2$ and $|T_\st|^2$ 
on $s$, together with the $K$-matrix approximation.}
\label{F9}
\end{figure}

In figs.\ref{F8} and \ref{F9} we display the isospin dependence of $\d_I$
and $|T_I|^2$.
It is worth noting that, by construction, $\d_\so$ passes through $90^0$ 
at $s=m_\k^2$ and, again, the $K$-matrix approximation is crude.
In the case of the $I=3/2$ channel, the free parameter in $\Rb_\st$ was fixed 
by imposing that the predicted phase shifts around $s=0.9$ GeV agree 
roughly with those given in Ref.\cite{Bachir}.
This gives rise to huge differences between full and $K$-matrix results.

\begin{figure}[h]
\hspace*{-25mm}
\includegraphics[width=0.6\columnwidth,angle=0]
{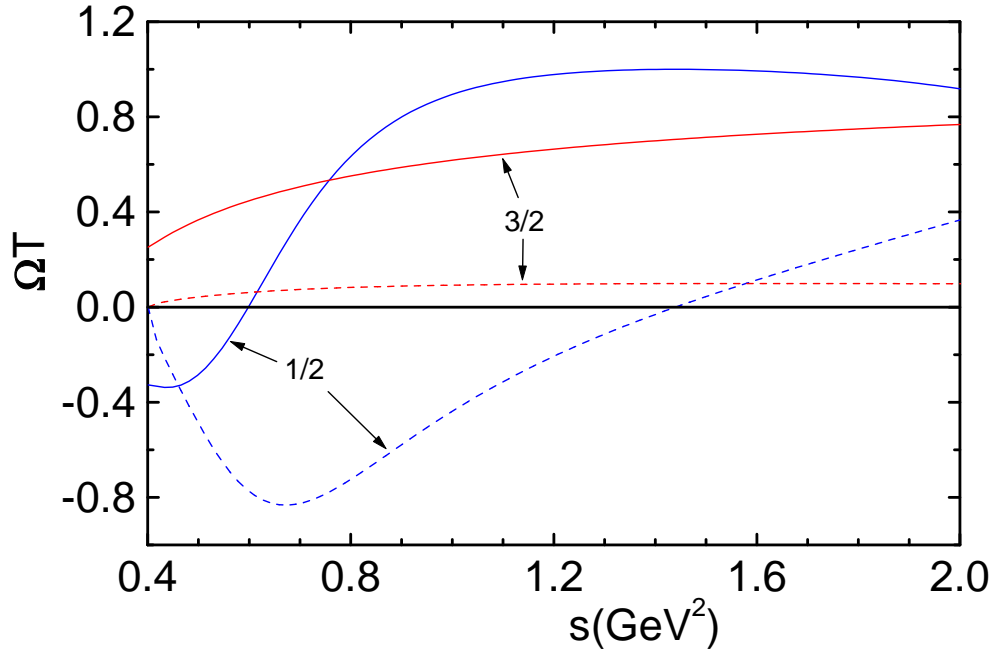}
\caption{(Color online) Real (continuous line) and imaginary (dashed line) 
components of the functions $\Ob_I \, T_I$.}
\label{F10}
\end{figure}

In the decay $D^+ \! \rar K^- \p^+ \p^+$, elastic amplitudes
$T_I$ contribute to final state interactions only.
They are always
accompanied by the two-body propagator $\Ob_I$, as indicated in fig.\ref{F2}
[please see also eqs.(\ref{7.2}-\ref{7.5})]. 
Real and imaginary parts of the products $\Ob_I\,T_I$ are displayed 
in fig.\ref{F10}.
In order to clarify the meaning of this figure, 
one notes that, in the $K$-matrix approximation, 
$\Ob \rar - i \, \rho/(16\p)$,  
and $\lb \Re\; \Ob\,T, \Im \; \Ob\,T \rb$ is given by 
$\lb \rho/(16\p) \lp \Im\; T, - \Re\; T\rp \rb$.
When the real part of $\Ob$ is turned on, a shift in the curves occur 
and the phase of $\Ob_I \, T_I$ becomes $\d_I \!+ \o_I$.
The magnitudes of the $\o_I$ may be inferred by noting that, in the $K$-matix
case, $\Re\; \Ob\,T =\Im\; T=0$ at threshold.
Loop phases are, therefore, explicit ingredients of the decay amplitude.
With future purposes in mind, one notes that the condition
$-1 \leq \Ob_\so \, T_\so \leq 1$ holds for both the real and imaginary 
components of this quantity.

\section{FSI: production amplitude}
\label{production}

\begin{figure}[h]
\begin{center}
\includegraphics[width=0.8\columnwidth,angle=0]{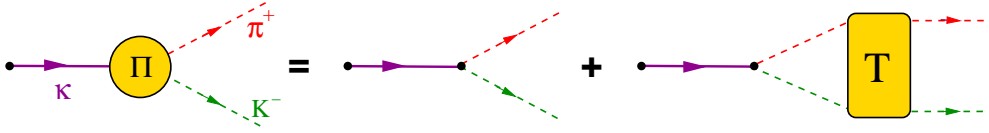}
\caption{(Color online) Resonance (full line) propagation and decay into $\p^+ K^-$ 
(dashed lines); $T$ is the unitary $I=1/2$ scattering amplitude.}
\label{F11}
\end{center}      
\end{figure}

We consider the possibility, shown in fig.\ref{F2}f, that the 
resonance can be directly produced at the weak vertex.
The interpolation between the decay and the observed $\p^+ K^-$
final state is described by the subset of diagrams shown in fig.\ref{F11}, 
denoted by $\Pi_\so$ and referred to as {\it production subamplitude}.
It involves both the bare $\k$ propagator and the unitarized elastic 
scattering amplitude $T_\so$, given by eq.(\ref{5.8}).
Reading the diagrams, one finds 
\beq
i\,\Pi_\so = \frac{\sqrt{3}/F^2}{s - m_\k^2} 
\lb c_d \, (s \sm M_\p^2 \sm M_K^2) + c_m \, (4\,M_K^2 \sp 5\,M_\p^2)/6 \rb\;
\lb 1 - \Ob_\so \, T_\so \rb \;.
\label{6.1}
\eeq

\ni
This function is complex, owing to the factor
$[1 \sm \Ob_\so \, T_\so]$, and can be cast in various fully equivalent 
forms\cite{BR07}, namely
\bea
i\,\Pi_\so \!\!&=& \!\! g\; \frac{\cos \d_\so}{m_\k^2 - s} \;
\lb 1 + \frac{\tan \d_\so}{\tan \o_\so} \rb \, e^{i \d_\so} 
\nn\\[2mm]
\!\!&=& \!\! g \; \frac{\cos \d_\so}{\cM_\k^2 - s} \; e^{i \d_\so} \;
=  g \;\frac{\sin \d_\so}{m_\k\, \Gamma_\k} \;e^{i \d_\so} \;,
\label{6.1b}\\[2mm]
g \!\!&=& \!\!  -(\sqrt{3}/F^2) \;  
\lb c_d \, (s \sm M_\p^2 \sm M_K^2) + c_m \, (4\,M_K^2 \sp 5\,M_\p^2)/6 \rb \;.
\label{6.2}
\eea

\begin{figure}[h]
\begin{center}
\hspace*{-40mm}
\includegraphics[width=0.60\columnwidth,angle=0]{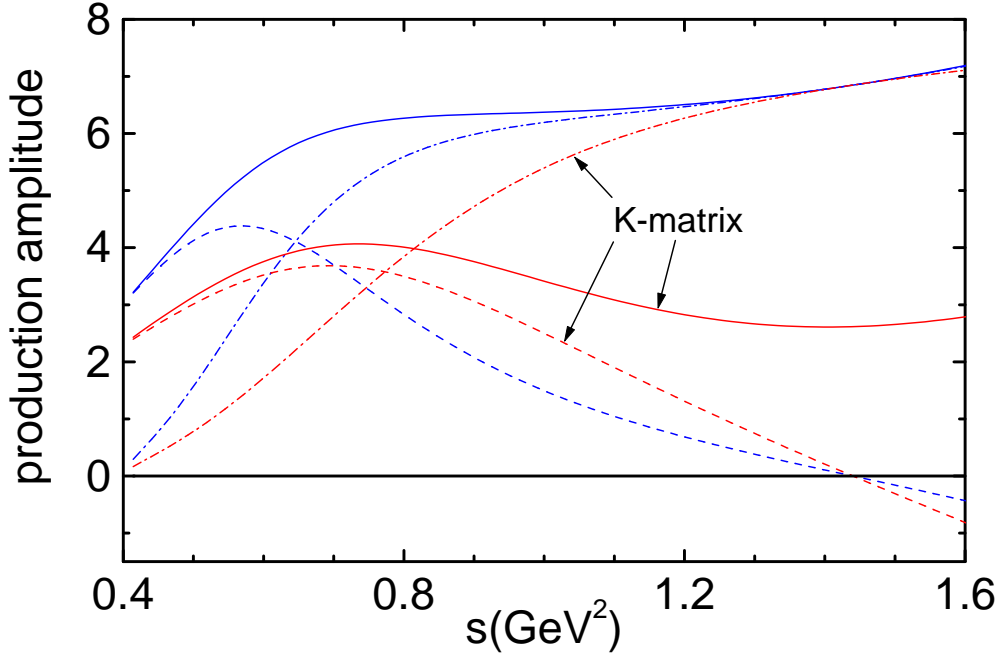}
\caption{(Color online) Modulus (full line), real and imaginary 
components (dashed and dot-dashed lines) of the function $i\,\Pi_\so(s)$,
eqs.(\ref{6.1}-\ref{6.2}), together with the $K$-matrix approximation.}
\label{F12}
\end{center}      
\end{figure}

Results have very little model dependence and show that, 
as predicted by Watson's theorem, the phase in the production 
amplitude is $\d_\so$, the same as in the elastic process. 
On the other hand, the magnitude of $\Pi_\so$ is determined by 
both $\d_\so$ and $\o_\so$.
It is important to stress that the full equivalence among these
different forms holds only  for the running mass and width 
given by eqs.(\ref{5.9}) and (\ref{5.10}).
If other forms for these functions are employed, consistency is lost
and results are no longer under control.
As in the elastic case, the function $\Pi_\so$
contains the $K$-matrix approximation as a particular case.
In Fig.\ref{F12} we show the behavior of the function $i\Pi_\so(s)$,
together with the corresponding $K$-matrix approximation, and
notes that the differences between both sets of results are important. 
The figure also indicates that bumps in the former are more pronounced.

\section{Amplitude for $D^+ \rar \p^+ \p^+ K^-$}

We display here individual contributions to the $D^+$ decay width
from the diagrams of fig.\ref{F2} , 
using both the scattering and production amplitudes derived previously.
They are covariant and expressed in terms of the invariant masses 
$\m_{ij}$, defined in appendix A.
The sum $\cA= [\cA_a + \cdots +\cA_f]$ is designed to replace the
background plus $S$-wave factor
$[a_0 \; e^{i\phi_0}\; \cA_0 + a_1^S \; e^{i \phi_1^S}\; \cA_1^S]$
mentioned in the introduction.

\ni
{\bf diagram (a):}
\beq
\cA_a(\mu_{\p \p}^2) = \frac{1}{3\sqrt{2}}\; [\d_a] \; G_F\cos^2\theta_C \;
\m_{\p\p}^2 \;;
\label{7.1}
\eeq

\ni
{\bf diagram (b):}
\bea
\cA_b(\mu_{K\p'}^2) \!\!&=&\!\!  -\,\frac{1}{3\sqrt{2}}\; 
[\d_a] \; G_F\cos^2\theta_C \;
\lb (P\cd q \sp M_\p^2) -\frac{(P\cd q \sm M_\p^2) \, (M_K^2 \sm M_\p^2)}
{M_D^2 \sp M_\p^2 \sm 2\,P \cd q} \rb 
\nn\\[2mm]
&\times& \!\! \lb \frac{2}{3}\; \Ob_\so \; T_\so(\mu_{K \p'}^2) 
+ \frac{1}{3}\; \Ob_\st \; T_\st(\mu_{K \p'}^2) \rb \;; 
\label{7.2}
\eea
\\
\ni
{\bf diagrams (c$+$d):}
\bea
\cA_{c + d}(\mu_{K\p}^2) &=& -\,\frac{1}{3\sqrt{2}}\; 
[\d_a] \; G_F\cos^2\theta_C \;
\lb (P\cd q' \sp M_\p^2) -\frac{(P\cd q' \sm M_\p^2) \, (M_K^2 \sm M_\p^2)}
{M_D^2 \sp M_\p^2 \sm 2\,P \cd q'} \rb \nn\\[2mm]
&\times&\Ob_\so \; T_\so(\mu_{K\p}^2) \;; 
\label{7.3}
\eea

\ni
{\bf diagram (e):}
\bea
\cA_e(\mu_{K\p}^2) \!\!&=&\!\!  -\,\frac{\sqrt{2}}{3}\; 
[\d_b] \; G_F\cos^2\theta_C \;
\lb (M_D^2 \sm 3\, P\cd q') 
-\frac{(M_D^2 \sm P\cd q') \, (M_K^2 \sm M_\p^2)}
{M_D^2 \sp M_\p^2 \sm 2\,P \cd q'} \rb 
\nn\\[2mm]
&\times& \!\! \lb \Ob_\so \; T_\so(\mu_{K\p}^2) 
- \Ob_\st \; T_\st(\mu_{K\p}^2) \rb \;;
\label{7.4}
\eea

\ni
{\bf diagram (f):}
\bea
\cA_f(\mu_{K \p}^2) \!\!&=&\!\!  - 4 \sqrt{3} \, 
[\d_c] \; G_F\cos^2\theta_C \;  
\frac{P\cd q'}{\m_{K\p}^2 - m_\k^2} \; [c_d/F^2]
\nn\\[2mm]
&\times& \!\!
\lb c_d \, (\mu_{K \p}^2 \sm M_\p^2 \sm M_K^2) 
+ c_m \, (4\,M_K^2 \sp 5\,M_\p^2)/6 \rb\;
\lb 1 - \Ob_\so \, T_\so(\mu_{K\p}^2) \rb \;. 
\label{7.5}
\eea

The process $D^+ \rar K^- \p^+ \p^+$ is Cabibbo allowed and, as expected,
amplitudes share the factor $G_F\cos^2\theta_C$.
As far as phases are concerned, one finds three kinds of structures.
The amplitude $\cA_a$ comes from a tree diagram and is necessarily real.
The phases of the amplitudes $\cA_b$, $\cA_{c+d}$ and $\cA_e$, 
on the other hand, are contained in the products $\Ob_I \, T_I$ and 
given by $(\d_I+\o_I)$.
Finally, as discussed in \cite{BR07}, the phase of $\cA_f$ is $\d_\so$,
the same of free scattering.

\begin{figure}[h] 
\hspace*{-28mm}
\includegraphics[width=0.38\columnwidth,angle=0]{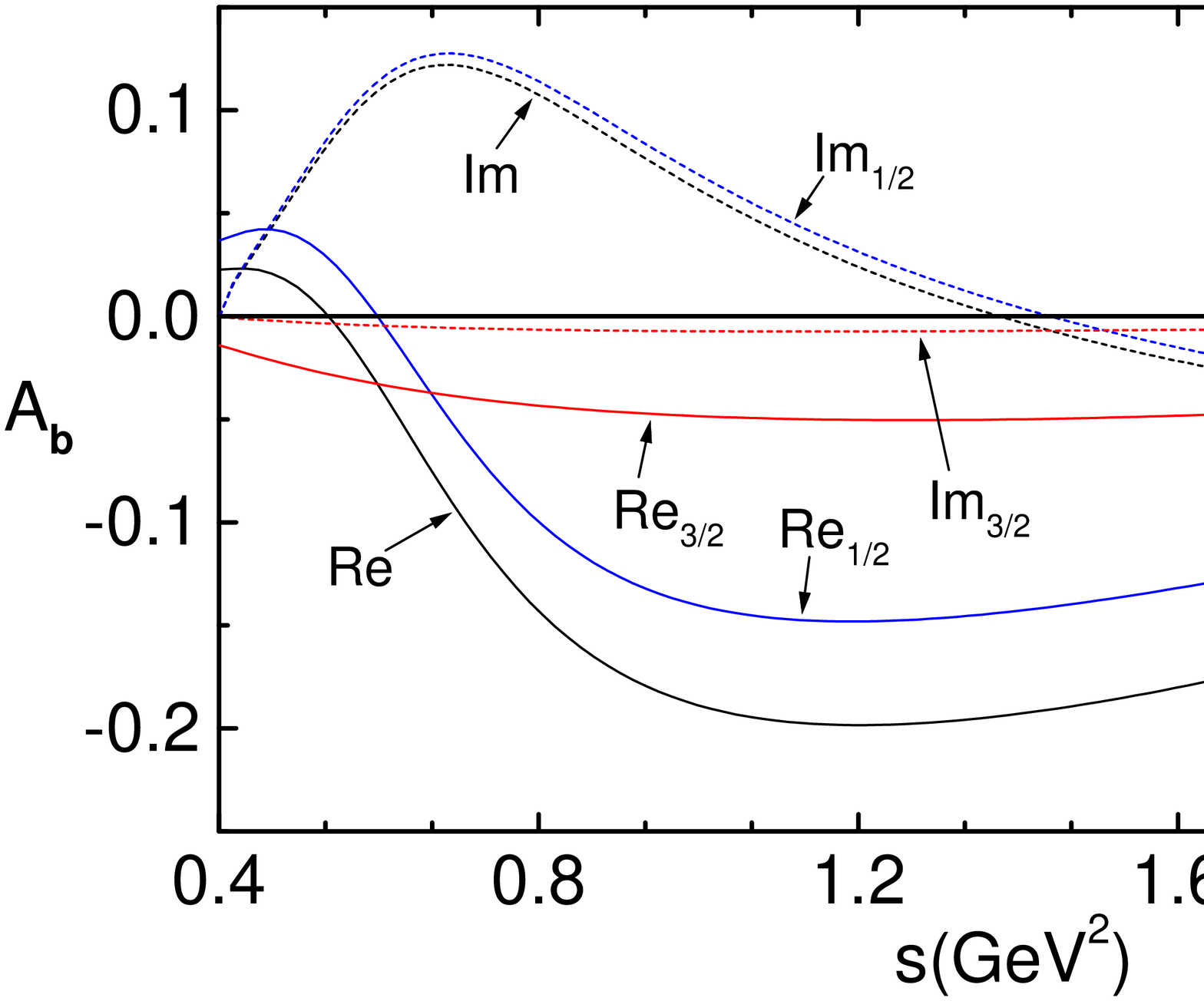}
\hspace*{20mm}
\includegraphics[width=0.38\columnwidth,angle=0]{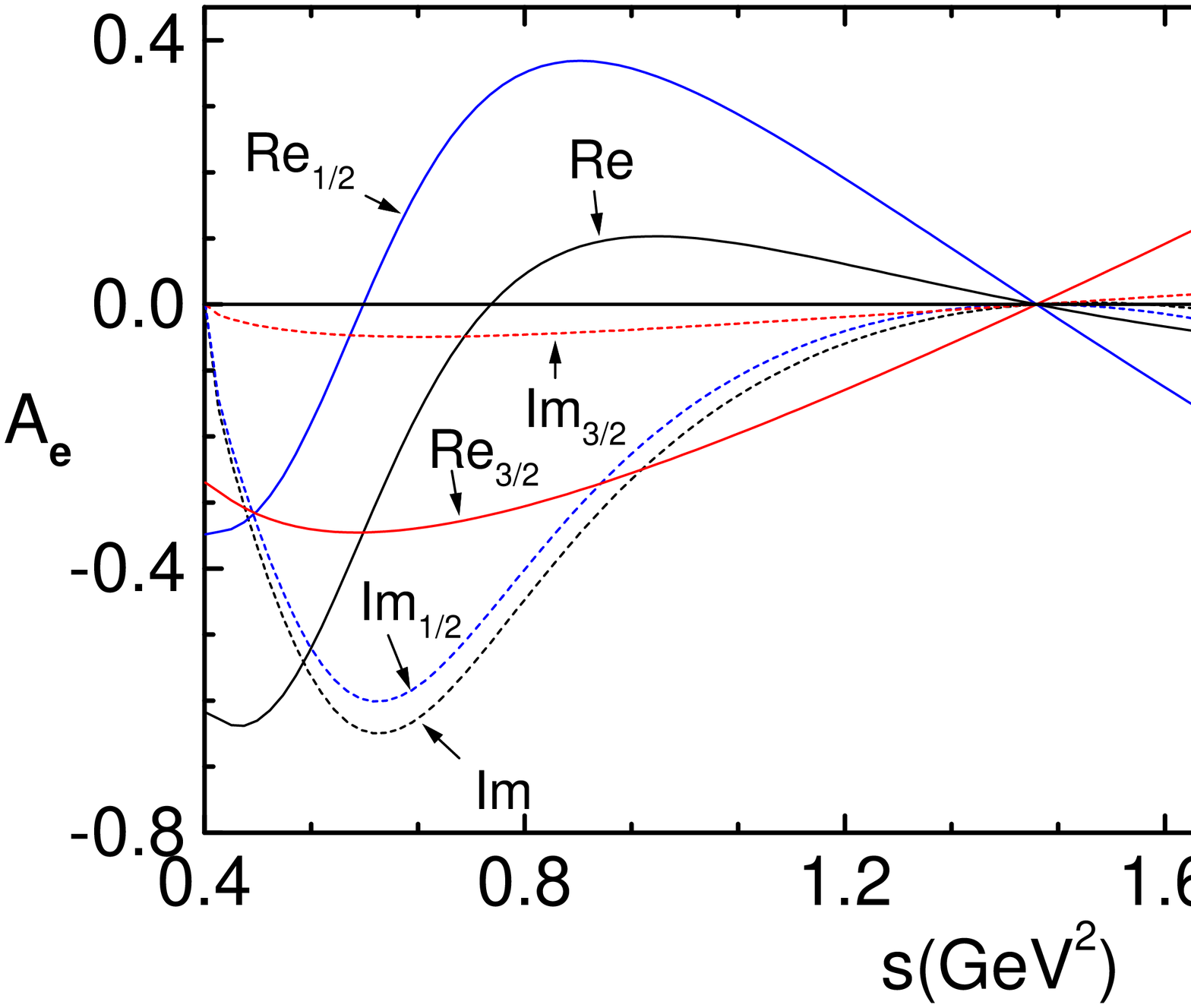}
\caption{(Color online) Full (black), $I=1/2$ (blue) and $I=3/2$ (red)
contributions to the real (continuous line) and imaginary (dashed line)
components of the amplitudes $\cA_b$ and $\cA_e$;
the vertical scale has to be multiplied by the weak factor $G_F\cos^2\theta_C$.}
\label{F13}
\end{figure}

In the evaluation of Dalitz plots, the amplitude $\cA$ has to be symmetrized
with respect to the variables $\m_{K\p}^2$ and $\m_{K\p'}^2$, since
outgoing pions are identical.
Allowed values for these invariant masses lie in the interval
$0.40\,$GeV$^2 \leq \m_{K\p}^2,\m_{K\p'}^2 \leq 2.99\,$GeV$^2$,
whereas the diagonal of Dalitz plot corresponds to
$0.94\,$GeV$^2 \leq \m_{K\p}^2=\m_{K\p'}^2 \leq 1.85\,$GeV$^2$.
As discussed in section \ref{scattering}, we assume our results 
to be valid for $s \leq 1.6\,$GeV$^2$ in the two-body channel.

Isospin $3/2$ contributions are present in $\cA_b$ and $\cA_e$ only.
Their dependence on isospin is displayed in 
figs.\ref{F13}, where it is possible to see that, 
for $I=3/2$, just the real part is relevant.

\begin{figure}[h] 
\hspace*{-35mm}
\includegraphics[width=0.7\columnwidth,angle=0]{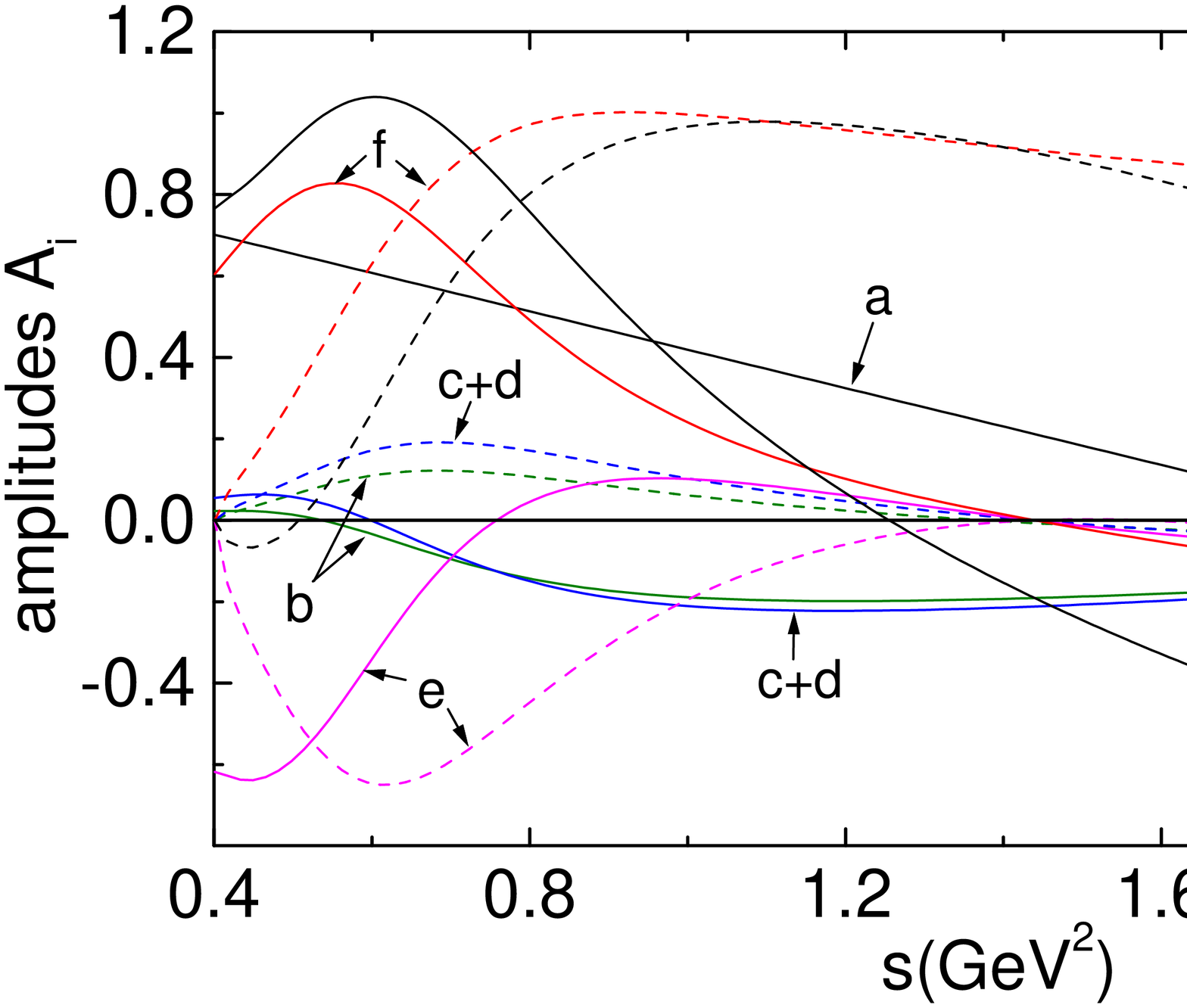}
\caption{(Color online) Full (black) and partial contributions to the 
real (continuous line) and imaginary (dashed line)
components of the amplitudes $\cA_i$;
the vertical scale has to be multiplied by the weak factor $G_F\cos^2\theta_C$.}
\label{F14}
\end{figure}

Real and imaginary components of the amplitudes $\cA_i$, as functions 
of $\m_{K\p}^2$, are shown in fig.\ref{F14}, for the choice
$m_\k=1.2\;$GeV and $\d_a=\d_b=\d_c=1$.
As the amplitude $\cA_a$ depends on $\m_{\p\p}^2$, we rewrote it as
\bea
&& \cA_a(\mu_{\p \p}^2) = [\bar{\cA}_a(\m_{k\p}^2) +
\bar{\cA}_a(\m_{k\p'}^2)]/2\;,
\nn\\[2mm]
&&
\bar{\cA}_a(x)=  
\frac{1}{3\sqrt{2}}\; [\d_a] \; G_F\cos^2\theta_C \;
(M_D^2 \sp 2M_\p^2 \sp M_K^2 -x)  
\label{7.6}
\eea
and just $\bar{\cA_a}(\m_{K\p}^2)$ was included in the figure.
All contributions have comparable magnitudes in this range, 
with a dominance of diagrams (a) and (f) in fig.\ref{F2}, which represent
the non-resonant background and the direct production of the resonance at 
the weak vertex.
We note, however, that the latter is rather sensitive to the resonance 
coupling constants $c_d$ and $c_m$ and recall that the values adopted here
are just illustrative.

\section{Dalitz plots}

In order to produce a feeling for the $\cA_i$ given by 
eqs.(\ref{7.1}-\ref{7.5}), we display here their predictions
for Dalitz plots. 
The plotted quantity is $|\cA_i(\m_{K\p}) \sp \cA_i(\m_{K\p'})|^2$
and individual contributions correspond to diagrams in fig.\ref{F2}.
Our description for the amplitudes $\cA_i$ is valid for invariant 
masses below $1.6\,$GeV$^2$.
As, in the plots, this condition must hold simultaneously for both amplitudes,
a reliable region around the lower side of the diagonal is selected.

\begin{figure}[h]
\begin{center}
\includegraphics[width=0.60\columnwidth,angle=0]{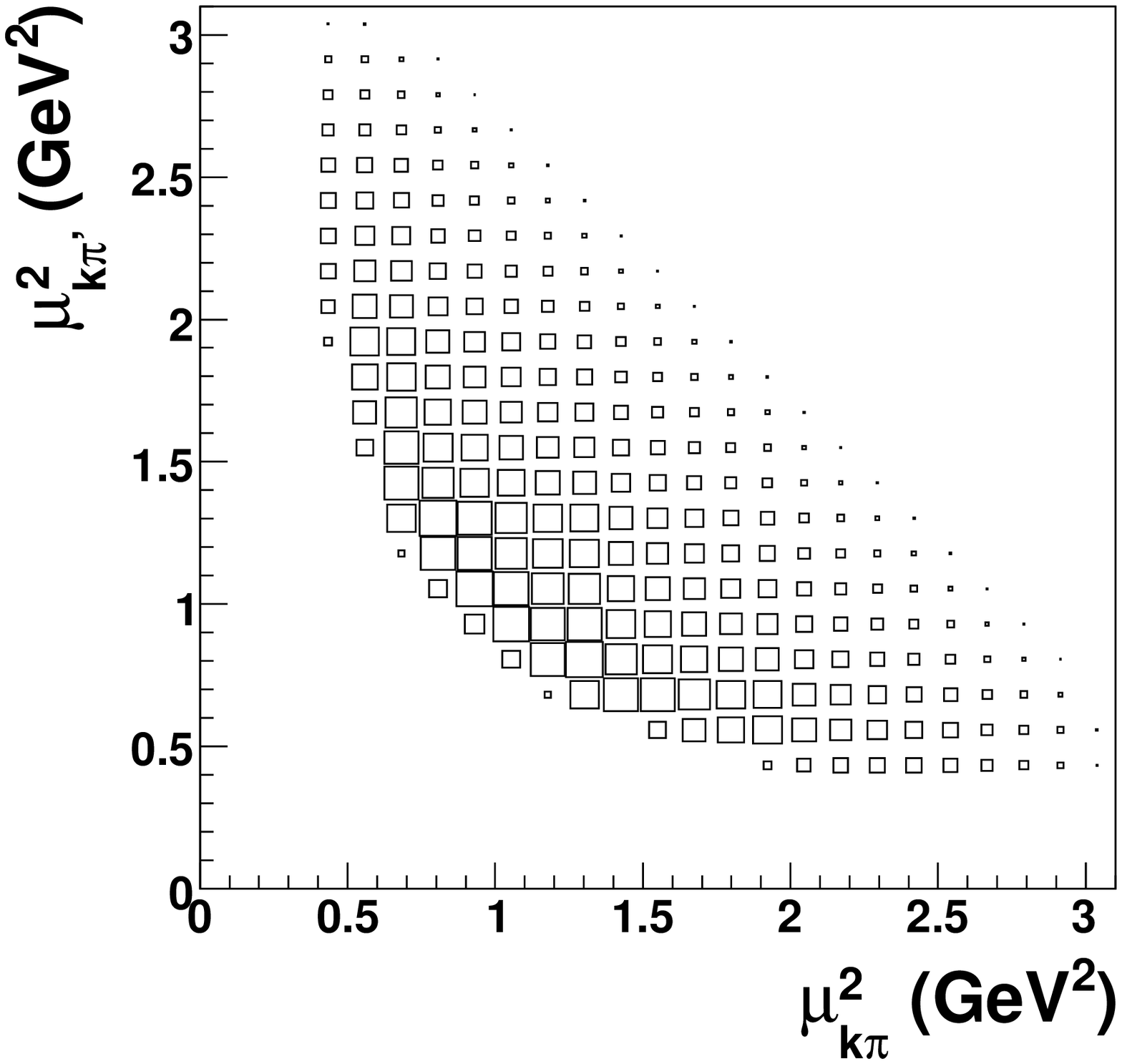}
\caption{Contribution from diagram (a), which represents the 
non-resonant background.} 
\label{F15} 
\end{center} 
\end{figure}

\begin{figure}[h]
\begin{center}
\includegraphics[width=0.40\columnwidth,angle=0]{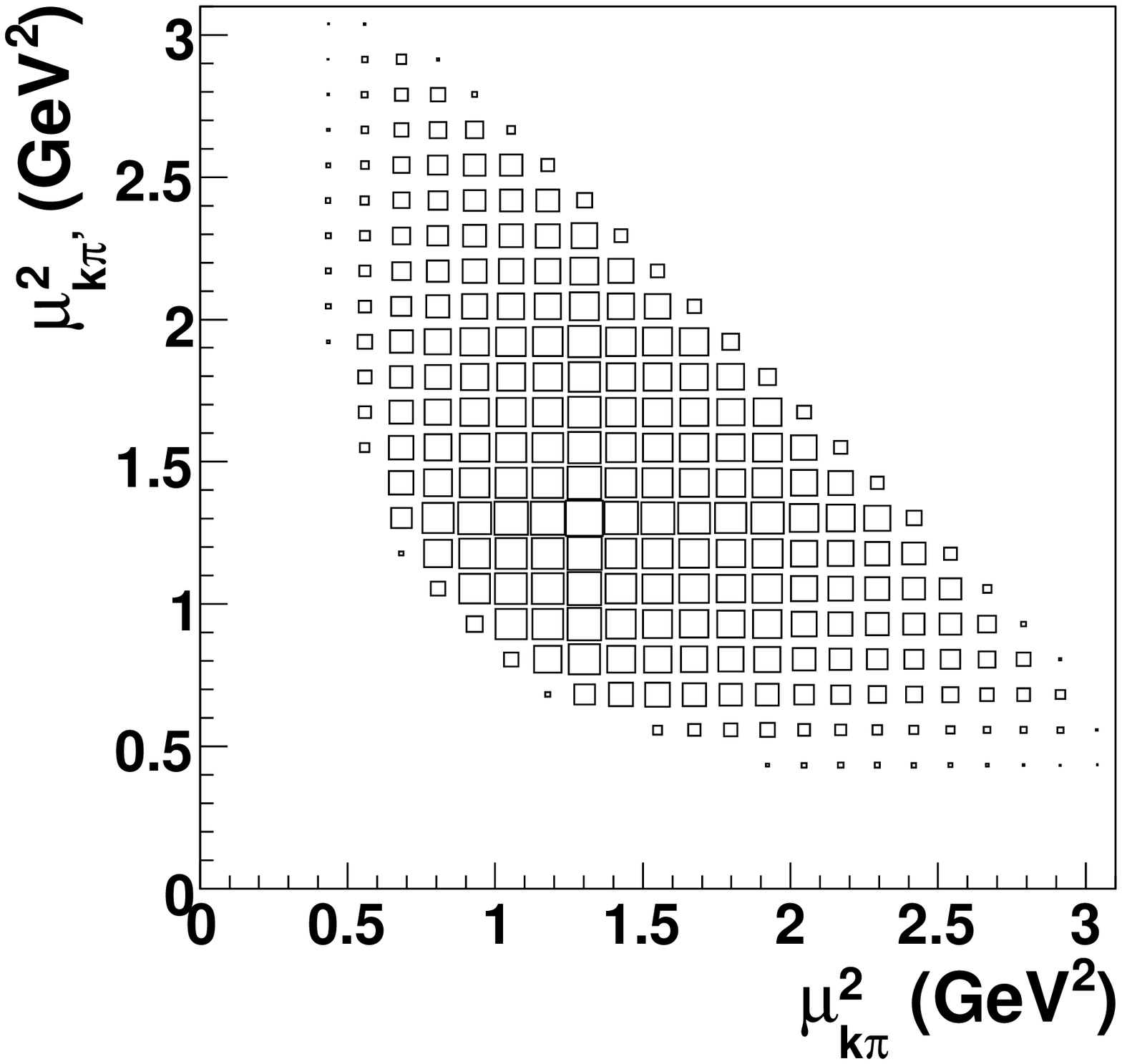}
\hspace*{15mm}
\includegraphics[width=0.40\columnwidth,angle=0]{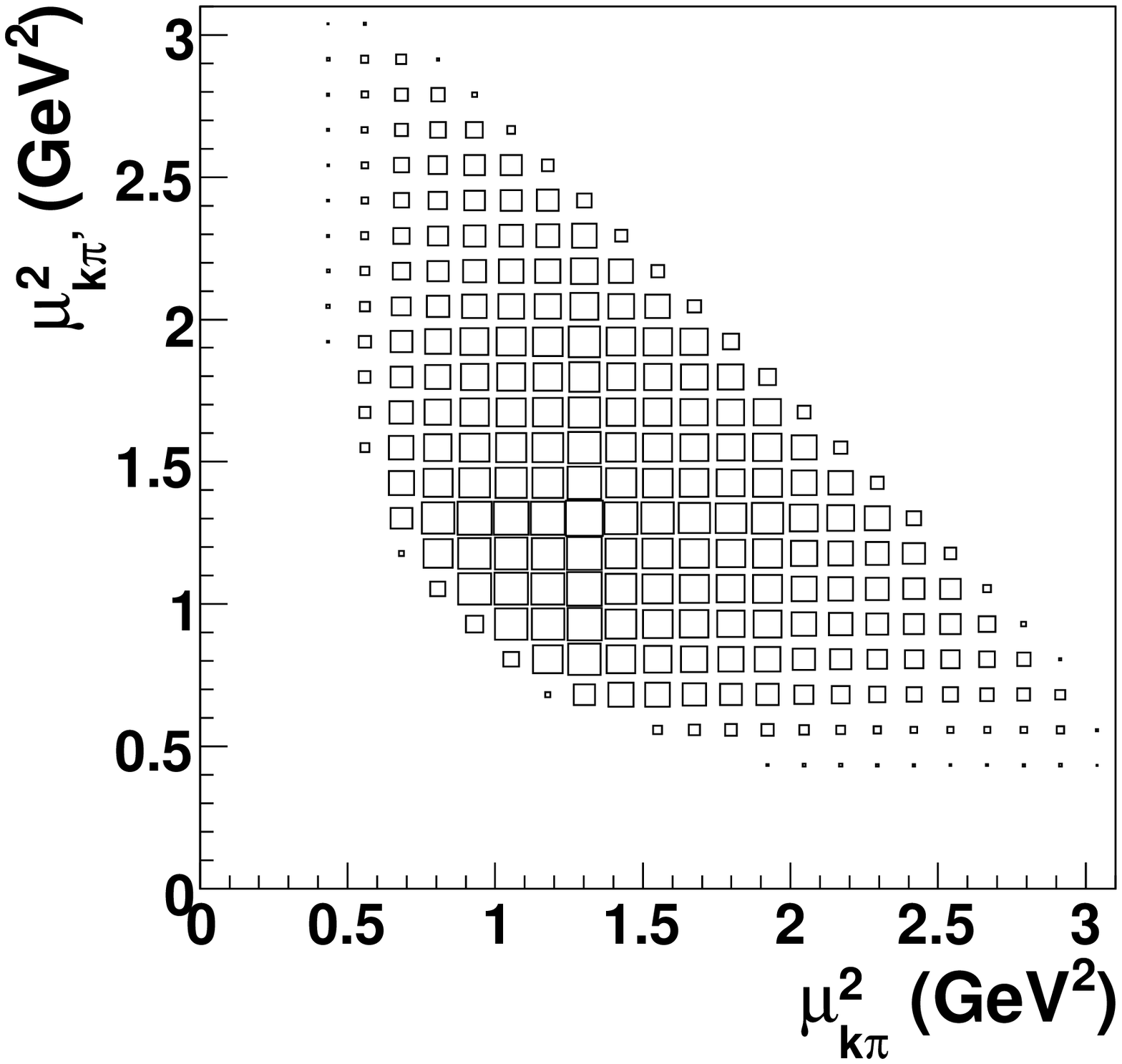}\\[5mm]
\includegraphics[width=0.40\columnwidth,angle=0]{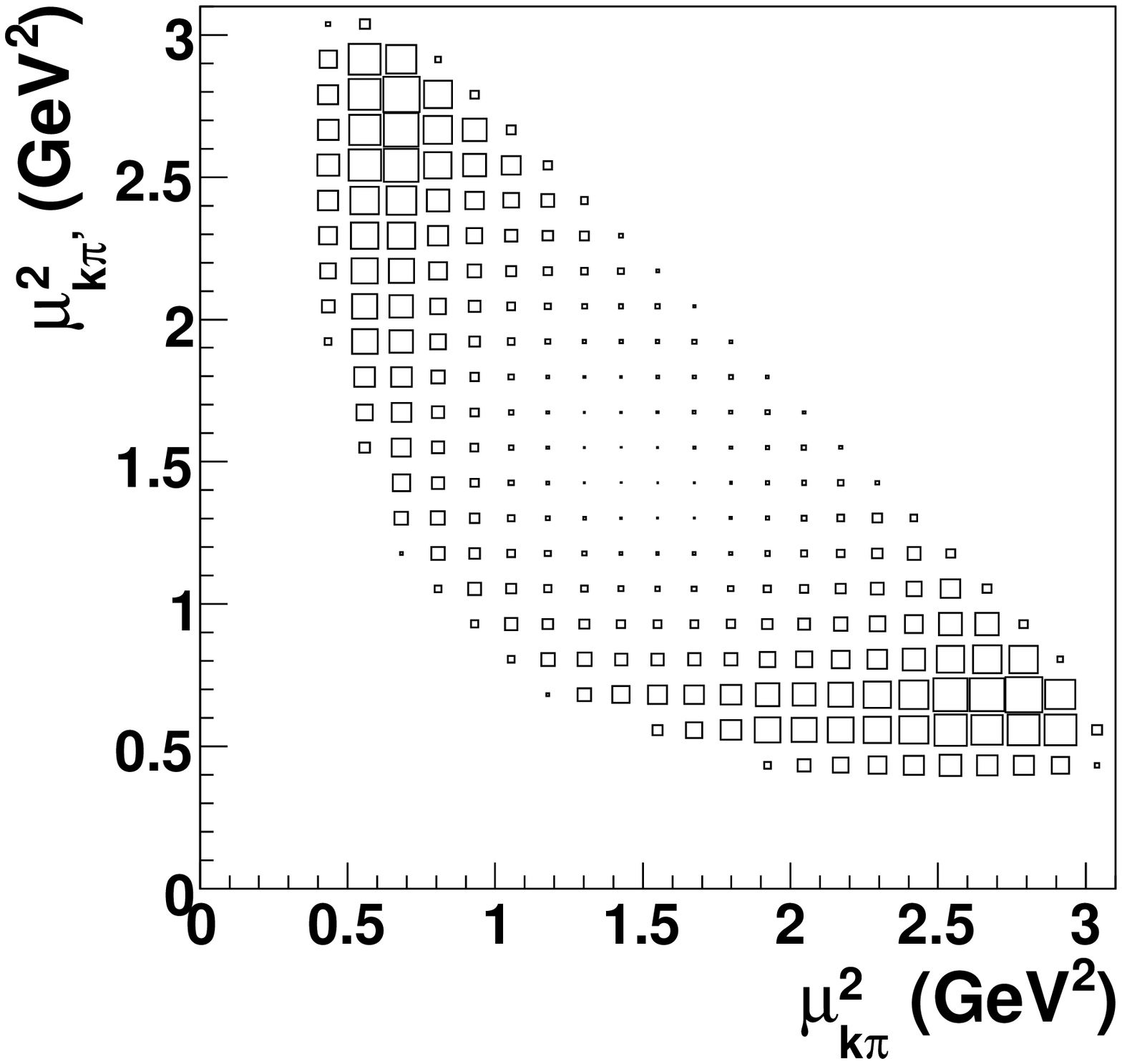}
\hspace*{15mm}
\includegraphics[width=0.40\columnwidth,angle=0]{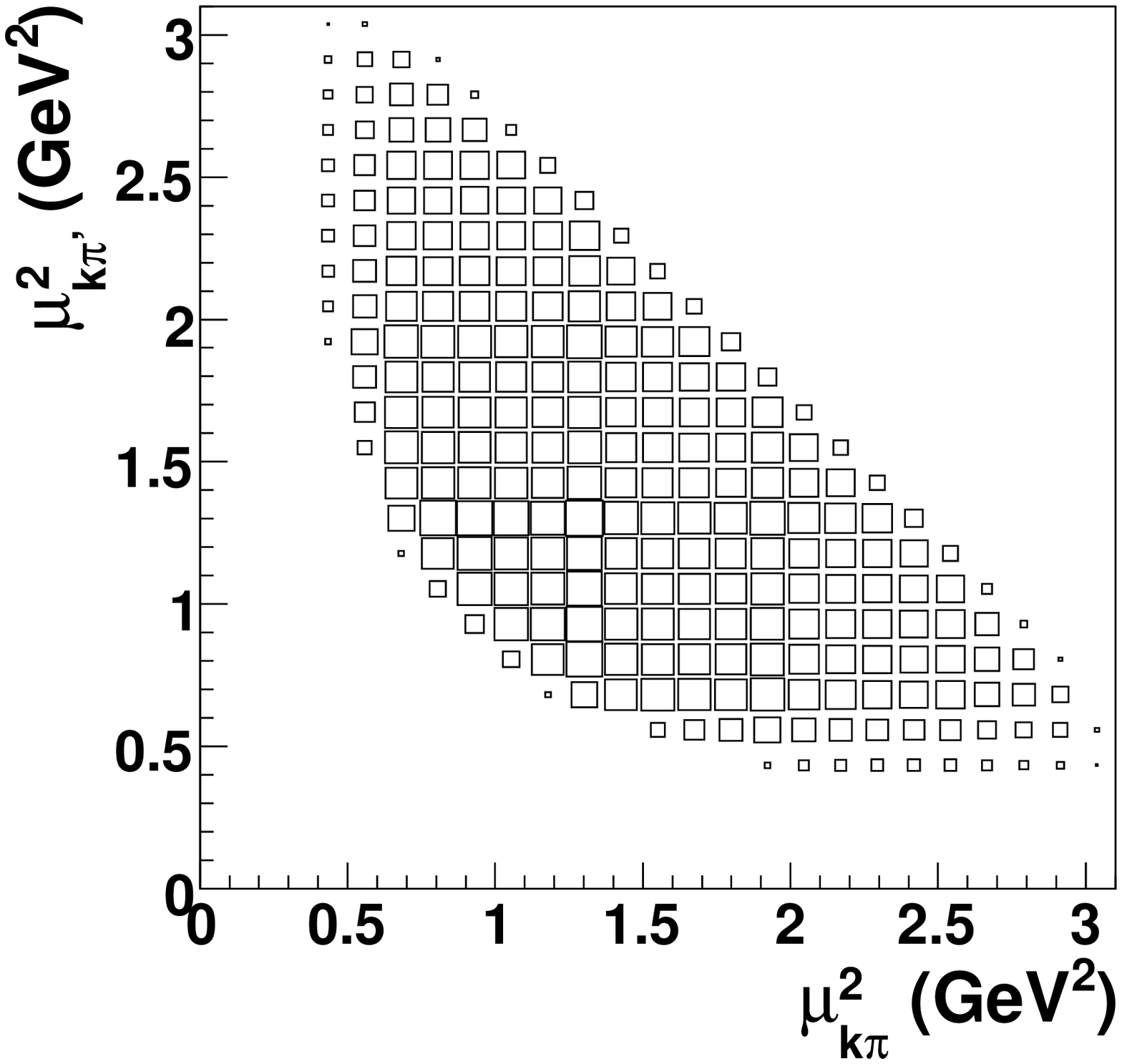}
\caption{Contributions from processes involving FSIs from the following
diagrams: (b) top-left; (c+d) top-right;
(e) bottom-left; (f) bottom-right.} 
\label{F16} 
\end{center} 
\end{figure}

\begin{figure}[h]
\begin{center}
\includegraphics[width=0.60\columnwidth,angle=0]{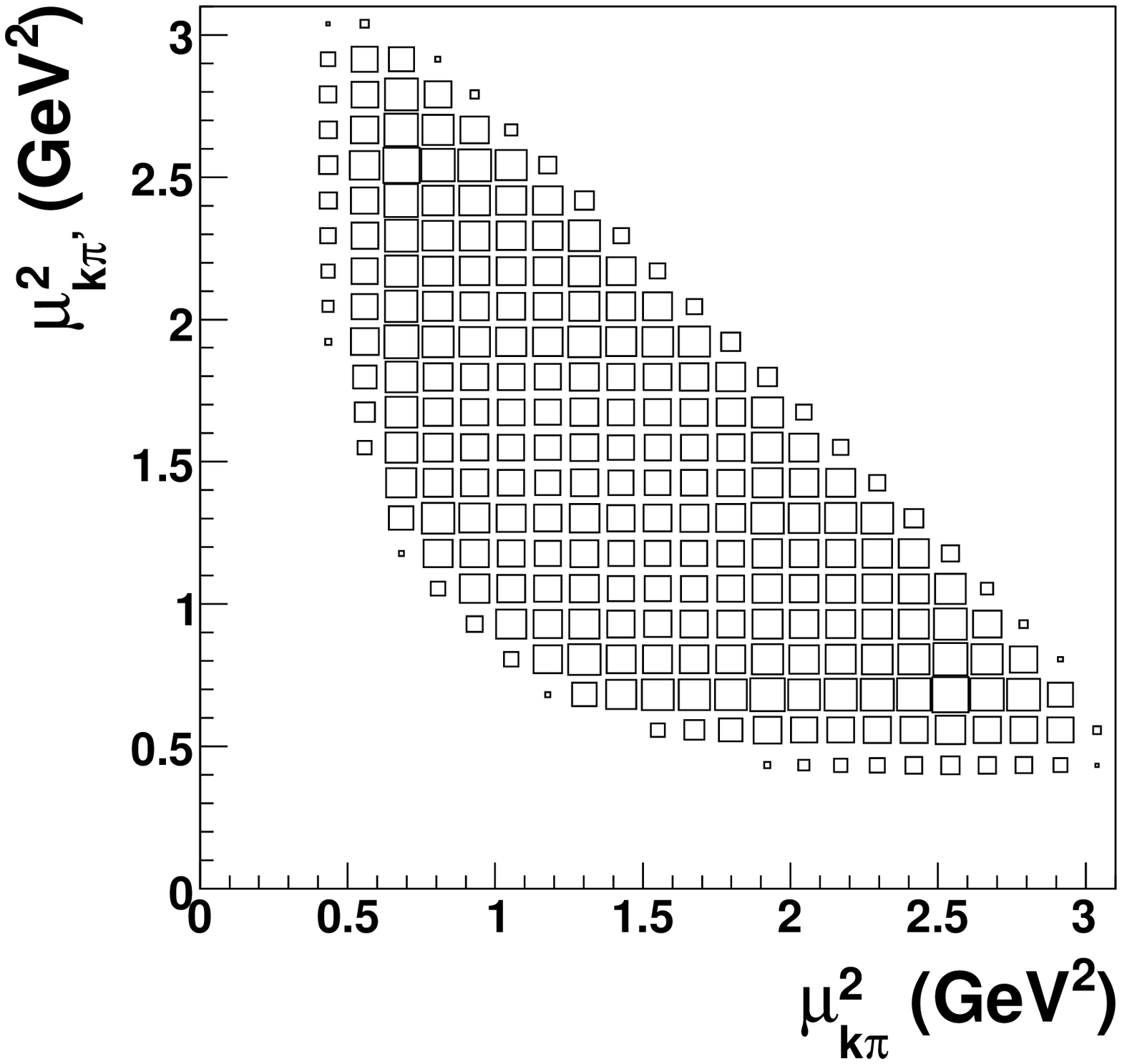}
\caption{Contributions from diagrams (a $+ \cdots$ f).} 
\label{F17} 
\end{center} 
\end{figure}

Diagram (2.a) gives rise to fig.\ref{F15}.
It describes  the non-resonating background and one learns that 
it is not evenly distributed along the plot, 
as sometimes assumed in the literature. 
We show, in fig.\ref{F16}, individual contributions from diagrams (2.b-f),
which involve final state interactions.
Amplitudes $\cA_b$ and $\cA_{c+d}$ share same weak vertices, but
the latter is based on just the isospin $1/2$ kernel, whereas the former
contains an admixture of isospins.
However, their Dalitz plots are very similar, indicating that isospin
$3/2$ contributions are small.
The plot corresponding to $\cA_e$ is very different from the other ones
because its weak vertices are of the vector type and the $W$ is coupled 
to two mesons.
The direct production of the resonance at the weak vertex is associated with
$\cA_f$ and produces a plot with a rather broad peak around $1.3\,$GeV$^2$. 
Finally, the sum $[\cA_a + \cdots +\cA_f]$ is given in fig.\ref{F17},
where a typical interference pattern can be noted.

\section{summary and conclusions}
\label{sum}

\begin{figure}[h]
\begin{center}
\hspace*{-35mm}
\includegraphics[width=.7\columnwidth,angle=0]{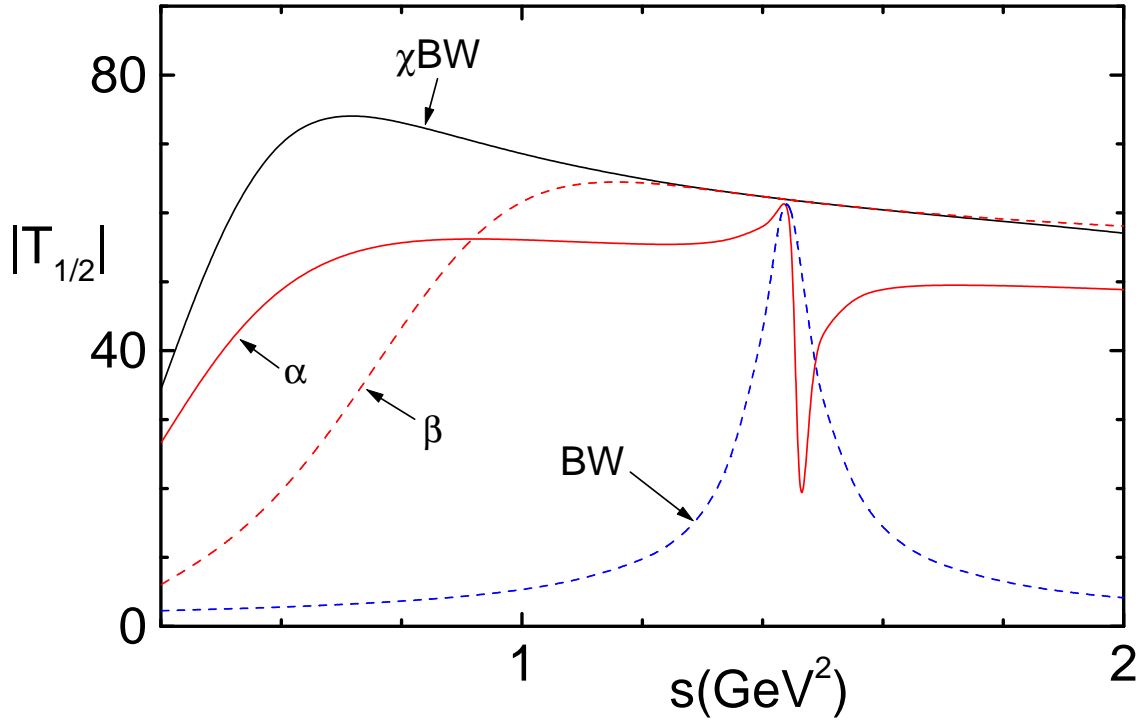}
\caption{(Color online)Relationship between a usual Breit-Wigner 
$(BW$, dashed line)
and its chiral generalization $(\chi BW$, continuous line);
the other curves correspond to the choices 
$\a:\,(\a=1, c_d=0)$ and $\beta:\, (\a=0, c_d=0.032\,$GeV)
in eq.(\ref{9.2}).} 
\label{F18} 
\end{center} 
\end{figure}

The low-energy components of the amplitude $D^+ \!\! \rar K^- \p^+ \p^+$ 
are studied in the framework of a rather conservative $SU(3)\times SU(3)$ 
chiral effective theory and special attention is paid to the resonance $\k$. 
For practical reasons, the derivation of weak vertices is performed 
using the group $SU(4)$, but without any commitment with the 
corresponding symmetry.
In dealing with final state interactions, proper three-meson processes 
are neglected and we remain within the quasi-two body approximation.
Our main results are summarized by eqs.(\ref{7.1}-\ref{7.5}), which represent
individual contributions from the diagrams in fig.\ref{F2}.
At low-energies, the decay amplitude is represented by 
$\cA=[\cA_a + \cdots +\cA_f]$ and symmetrization with respect to 
final pions is required.
Conclusions are presented in the sequence.\\[1mm]
{\bf 1. degrees of freedom:}
The amplitudes $\cA_i$ contain both fixed and adjustable parameters.
The former class encompasses pseudoscalar masses $M_i$, their decay 
constant $F=F_\p=0.093\,$GeV and the weak constants 
$G_F=1.166\times 10^{-5}\,$GeV$^{-2}$ and 
$\cos\theta_C=0.9745$ \cite{PDG}. 
Adjustable parameters involve resonance masses and coupling constants.
In principle, three scalar states should be considered, but two of them 
appear just in the $t$-channel background.
As discussed in appendix \ref{background}, contributions from 
$t$- and $u$-channel interactions are very small and can be safely 
neglected. 
In this approximation, the only free parameters in our results
are $m_\k$, the $\k\p K$ coupling constants $c_d$ and $c_m$,
and three $SU(4)$ breaking factors $[\d_i]$.\\[1mm]
{\bf 2. background:} The non-resonating background is represented by $\cA_a$, 
which is a {\em real} function.
It important to note that this does not happen by chance.
The process shown in fig.\ref{F2}a is the simplest possible in $D^+$ 
decays and given by a tree-level diagram.
In field theory, tree diagrams are real and imaginary components are 
produced by loops.
Of course, it would be possible to dress the primary weak vertex
with mesonic loops, but this would amount to including higher order chiral
corrections. 
Therefore, at low-energies, the background is necessarily real and should 
not be represented by trial functions of the form 
$[a_0 \; e^{i\phi_0}\; \cA_0]$.
Inspecting figs. \ref{F14} and \ref{F15}, one learns that the magnitude
of $\cA_a$ is comparable to other contributions and that its distribution
over the Dalitz plot is not uniform, as sometimes assumed.\\[1mm]
{\bf 3. phases:} Our calculation begins with a lagrangian, which yields 
real vertices only.
Loops are introduced in a subsequent step, and only then 
amplitudes become complex.
This construction process is systematic and one has full control 
over all imaginary terms and understand clearly their dynamical origins.
Complex amplitudes are due to final state interactions 
and encoded into the functions $\Ob_I$ and $T_I$, representing respectively
two-meson propagators and elastic $K\p$ scattering. 
As one deals with two isospin channels, in principle, four phases need
to be considered. 
However, the $I=3/2$ channel is repulsive and the corresponding 
phases are small.
One is then left with just $\o_\so$ and $\d_\so$.
Both of them are present in the phases of $\cA_b$, $\cA_{c+d}$ and $\cA_e$,
which are identical and given by $(\d_\so + \o_\so)$.
Diagram \ref{F2}f represents the direct production of the $\k$-resonance
at the weak vertex and involves the subamplitude $\Pi_\so$, represented 
in fig.\ref{F11}.
As pointed out in ref.\cite{BR07}, this structure gives rise to the same
phase as in $T_\so$, namely $\d_\so$.
Finally, one notes that another phase is produced when 
the complex $I=1/2$ amplitudes are added to the almost real $I=3/2$
conterparts.
In summary, the low-energy decay amplitude contains several different 
energy-dependent phases and cannot be well represented by a trial function 
such as $[a_1^S \; e^{i \phi_1^S}\; \cA_1^S]$.\\[1mm]
{\bf 4. two-body isospin channels:} Our numerical results indicate that 
final state interactions are very important in $D^+ \! \rar K^- \p^+\p^+$.
They are present in five of the diagrams shown in fig.\ref{F2}.
Three of them involve just the isospin $1/2$ channel, whereas the other
two depend on both components. 
These cases were studied in fig.\ref{F13} and one notes that isospin $3/2$
is relatively important in $\cA_e$ only.
\\[1mm]
{\bf 5. $K$-matrix approximation:}
This problem is addressed in section \ref{scattering}.
The main advantage of the $K$-matrix is its simplicity.
Our expressions encompass the $K$-matrix approximation,
since it amounts to neglecting the real part of the 
two-loop propagator and setting $\o_I=\p/2$.
In the isospin $3/2$ channel, this procedure is disastrous.
For $I=1/2$, on the other hand, it gives rise to reasonable qualitative
predictions. 
One should bear in mind, however, that the phase $\o_\so$ influences 
final result in two different ways, since it both
helps shaping the $K\p$ amplitude and enters directly
into the expressions for $\cA_b, \cdots ,\cA_f$.\\[1mm]
{\bf 6. dynamics:} We assume our results to be reliable for $K\p$
invariant masses between $0.4\,$GeV$^2$ and $1.6\,$GeV$^2$.
The magnitudes of the $\cA_i$ are comparable in this range,
ss shown in fig.\ref{F14}.
There, it is also possible 
to see that the non-resonant background and the 
direct resonance production dominate.
However, numerical results are particularly sensitive to the coupling 
constant $c_d$ and, for the time being, we take this conclusion
as provisional.\\[1mm]
{\bf 7. Breit-Wigner expressions:} Our results involve  
functions which are akin to the usual Breit-Wigner ones.
However, as we discuss in the sequence, differences between them are very 
important. 
These functions are hidden in the two-body amplitudes $T_\so$,
present in diagrams (\ref{F2} b, ... ,e) , and
also contribute to the $\Pi_\so$ in (f).
Eq.(\ref{5.8}) can be rewritten as 
\beq
T_\so = \frac{\g^2}
{[(m_\k^2 -s)+\g^2 \,\Rb_\so]   -i\,[\g^2 \,\rho/(16 \p)]}\;,
\label{9.1}
\eeq 
where $\Rb_\so$ represents off-shell effects in the two-meson propagator.
The function $\g^2$ is given by eq.(\ref{4.7}) and can be rewritten as
\bea
\g^2 \!\!&=&\!\! 3\,h^2/F^2 
+ \lb \a \, \Lambda + c_d \,\beta \rb (m_\k^2 \sm s) \;,
\label{9.2}\\[2mm]
h \!\!&=&\!\! 
[c_d\; \lp m_\k^2 \sm M_\p^2 \sm M_K^2 \rp 
+ c_m \; \lp 4\,M_K^2 \sp 5\,M_\p^2 \rp /6 ] \;,
\nn\\[2mm]
\Lambda \!\!&=&\!\!  (1 / 4 F^2)\,
\lb \lp 4 -3\,\rho^2/2 \rp s - 4\, \lp M_\p^2 + M_K^2\rp \rb \;,
\nn\\[2mm]
\beta \!\!&=&\!\! [3\,c_d\, (m_\k^2 \sm s) - 6\, h]/F^3 \;. 
\eea
In this expression, $h^2$ is a $\cO(q^4)$ effective coupling constant, 
$\Lambda$ is the $\cO(q^2)$ leading term in the chiral amplitude
and $\beta$ is a $\cO(q^4)$ background.
A parameter $\a$ has been introduced, 
so that the leading chiral contribution could be turned on or off.
The usual Breit-Wigner expression can be recoverd from eq.(\ref{9.1}),
by going to the $K$-matrix approximation ($\Rb_\so \rar 0$) 
and by choosing $(\a=0, c_d=0)$.
This yields the curve $BW$ in fig.\ref{F18}, with its  
well known shape.
The reintroduction of $\Rb_\so$ produces no visible effects.
The choices $(\a=1, c_d=0)$ and $(\a=0, c_d=0.032\,$GeV) give rise 
respectively to curves $\a$ and $\beta$.
In both cases, one notes huge enhancements in the region 
$0.8\,$GeV$^2<s <m_\k^2$.
At threshold, on the other hand, the chiral hierarchy is respected,
since $\a \rar \cO(q^2)$ and $\beta \rar \cO(q^4)$. 
Finally, the curve $\chi BW$ is produced by $(\a=1, c_d=0.032\,$GeV$)$
and is directly related with those discussed in section \ref{scattering}.

\noindent
{\bf 8. chiral symmetry:}
The phase shifts predicted by eq.(\ref{9.1}) pass through $\p/2$ at $s=m_\k^2$. 
This feature defines the {\em nominal} mass of the resonance and
is completely independent of parameters adopted.
On the other hand, the implementation of chiral theorems requires that 
amplitudes be represented by polynomials which have well known values  
at threshold and grow close by.   
The form of the curve $\chi BW$ in fig.\ref{F18} corresponds to a compromise
between those two features.
It is important to note that this kind of behavior cannot be obtained by
adding polynomials to usual Breit-Wigner expressions.  
Rather, it derives directly from the unitarization of contact interactions 
added to resonance poles, as discussed in sections 
\ref{kernel} and  \ref{scattering}.
A study of pole movements induced by this procedure is in progress 
and will be presented elsewhere.

\begin{acknowledgments}
We thank Alberto Reis, Ana Am\'elia Bergamini Machado, Carla G\"obel and 
Ign\'acio Bediaga for discussions and information about empirical data.
This work is supported by FAPESP(Brazilian Agency).
The work by DRB  is supported in part by a FPI scholarship of the 
Ministerio de Educaci\'on y Ciencia under grant FPA2005-02211,
the EU Contract No.~MRTN-CT-2006-035482, ``FLAVIAnet'',
and the Spanish Consolider-Ingenio 2010 Programme CPAN (CSD2007-00042).
\end{acknowledgments}

\appendix
\section{kinematics}
\label{kinematics}

Initial and final momenta are uniformly
represented by capital and low-case letters.

\ni
{\bf two-body system:} In the description of the two-body reaction
$\p(Q)\, K(K) \rar \p(q)\, K(k)$, Mandelstam variables are defined by
$s = (Q \sp K)^2 = (q \sp k)^2$, 
$t = (Q \sm q)^2 = (K \sm k)^2$,
$u = (Q \sm k)^2 = (K \sm q)^2$
and satisfy the condition $s \sp t \sp u = 2\,(M_K^2 \sp M_\p^2)$.
In the center of mass, one has
\bea
&& t= -(s\,\rho^2/2)\;(1 \sm \cos\theta) \;,
\label{a.1}\\[2mm]
&& u= (M_K^2 \sm M_\p^2)^2/s - (s\,\rho^2/2)\;(1 \sp \cos\theta) \;,
\label{a.2}\\[2mm]
&& \rho=
\sqrt{|1 - 2\, (M_K^2 \sp M_\p^2)/s+ (M_K^2 \sm M_\p^2)^2/s^2|} \;.
\label{a.3}
\eea

In performing $S$-wave projections, one uses 
\beq
T_0 = (1/2) \;\int_0^\p d\cos\theta \; T 
\label{a.4}
\eeq
\ni
and finds
\bea
&& t\rar t_0 = -s\,\rho^2/2\;, \;\;\;\;\;\;\; 
u\rar u_0 = (M_K^2 \sm M_\p^2)^2/s - (s\,\rho^2/2)\;,
\label{a.5}\\[2mm]
&& \lb 1/(x \sm m^2) \rb_0 = -(1/ s\,\rho^2)\;
\ln[1 \sp s\,\rho^2/(m^2\sm x_0 - s\,\rho^2/2)]\;,
\label{a.6}
\eea

\ni
with $x = t, u$.

\ni
{\bf three-body system:} In the $D^+$ decay, momentum variables are defined
as $D^+(P) \rar \p^+(q) \, \p^+(q') \, K^-(k)$. 
The $W$ propagator splits diagrams into two parts and the momentum $(q)$
is associated with the pion in the same sector as the $D^+$. 
Invariant masses are described by 
\beq
\mu_{\p \p}^2 = (q \sp q')^2 , \;\;\;\;
\mu_{K \p}^2 = (q \sp k)^2 , \;\;\;\;
\mu_{K \p' }^2 = (q' \sp k)^2, 
\label{a.7}
\eeq

\ni
and one has $ \mu_{\p \p}^2 + \mu_{K \p}^2 + \mu_{K \p' }^2 
= M_D^2 + 2\, M_\p^2 + M_K^2$.
Scalar products are then given by 
$P \cdot q = (M_D^2 + M_\p^2 - \m_{K\p'}^2)/2$ and 
$ P \cdot q' = (M_D^2 + M_\p^2 - \m_{K \p}^2)/2 $ .

\section{two-meson propagator}
\label{propagator}

\begin{figure}[h] 
\includegraphics[width=0.3\columnwidth,angle=0]{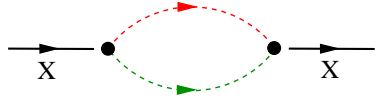}
\caption{(Color online) Two-meson propagator.}
\label{F19}
\end{figure}

The two-meson propagator is shown in fig.\ref{F19}, for a system with total 
momentum $X=Q \sp K= q \sp k$.
We also use the combination $\ell = (Q \sm K)/2$ and define
\beq
\lb I_{\p K};\, I_{\p K}^\m \rb = \int \frac{d^4 \ell}{(2 \p)^2} \;
\frac{\lb 1; \, \ell^\m \rb}
{[(\ell \sp X/2)^2 \sm M_\p^2]\;[(\ell \sm X/2)^2 \sm M_K^2]} \;.
\label{b.1} 
\eeq

In terms of Feynman parameters, these integrals read
\bea
&& \lb I_{\p K};\, I_{\p K}^\m \rb = \frac{i}{(4\p)^2}
\lb \Pi_{\p K}^{00}; \, \frac{X^\m}{2}\, 
\lp \Pi_{\p K}^{10} - \Pi_{\p K}^{01} \rp \rb \;,
\nn\\[2mm]
&& \Pi_{\p K}^{mn} = -\, \int_0^1 d\,a \; [-(1\sm a)^m]\,[-a^n]\,
\ln [D_{\p K}/\Lambda^2] + \cdots \;,
\nn\\[2mm]
&& D_{\p K} = (1\sm a)\,M_\p^2 + a\,M_K^2 - a(1 \sm a)\, X^2 \;.
\label{b.2} 
\eea

\ni
where the ellipsis indicate an infinite quantity associated with dimensional 
regularization.
Multiplying $I_{\p K}^\m$ by $X_\m$ in eq.(\ref{b.1}) and manipulating the 
integrand, one finds the following useful result
\beq
\lp \Pi_{\p K}^{10} - \Pi_{\p K}^{01} \rp =
\frac{M_\p^2 \sm M_K^2}{X^2} \, \Pi_{\p K}^{00}
+ \cdots \;,
\label{b.3}
\eeq

The two-meson propagator given by eq.(\ref{5.3}) is written 
as $\Omega = -(1/16 \p^2) \lb L(s) \sp \Lambda_\infty \rb$,
where the function $L(s)$ is explicitly given below and $\Lambda_\infty$
is an infinite constant that has to be removed by renormalization.
In this procedure, the function $\Omega$ is replaced by 
\beq 
\Ob_I = -(1/16 \p^2) \lb L(s) \sp c_I \rb \;,
\label{b.4}
\eeq

\ni
where the $c_I$ are constants which depend on the isospin channel.
They are chosen by tuning the predicted phase shifts $\d_I(s)$,
eq.(\ref{5.6}), to experimental results at a given point $s=s_I$
and one imposes $\d_I(s_I) \equiv \d_I^{\mathrm{exp}}(s_I)$.
When a resonance is present, a rather convenient choice\cite{BR07} 
for $s_I$ is the point at which the experimental phase is $\p/2$.

The function $L(s)$ entering eq.(\ref{b.4}) is given by
\bea
\bullet && \mathrm{for}\,\, s<(M_K - M_\p)^2 :
\nn\\[2mm] 
&& L(s) = - \rho\; \log \lb \frac{\s-1}{\s + 1}\rb 
-\eta \;,
\label{b.5}\\[2mm]
\bullet && \mathrm{for}\,\,(M_K - M_\p)^2< s< (M_K+M_\p)^2: 
\nn\\[2mm]
 && L(s) = \rho \; \lb \tan^{-1} \s - \pi/2  \rb 
 - \eta \;,
\label{b.6}\\[2mm]
\bullet && \mathrm{for}\,\, s>(M_K + M_\p)^2:
\nn\\[2mm]
&& L(s) = \rho(s) \log \lb \frac{1 - \s}{1 + \s}\rb 
- \eta + i \pi \rho\;,
\label{b.7}\\[2mm]
&&\;\;\;\;\; \s = \sqrt{|s \sm (M_K \sp M_\p)^2|/
|s \sm (M_K \sm M_\p)^2 | } \;,
\label{b.8}\\[2mm]
&& \;\;\;\;\;\eta = 2 - [(M_K^2 -M_\p^2)/s] \,\log( M_K/M_\p)] \;.
\label{b.9}
\eea

These results allow the renormalized two-loop propagator to be written as
\bea
\Ob_I &=& -\frac{1}{16\p^2} 
\lc \Re \lb L(s) - c_I \rb + i\, \Im  L(s)  \rc
\nn\\
&\equiv& \bar R_I(s) + i\, \theta[s \sm (M_K \sp M_\p)^2]\; I(s) \;.
\label{b.10}
\eea

\ni
The imaginary component is very simple and reads $I(s)=-\rho/(16\p)$.
We define a loop phase $\omega_I$ by
\beq
\tan \omega_I \equiv I/\bar R_I \;.
\label{b.11}
\eeq

For the channel $I=1/2$, the constant $c_{1/2}$ is chosen so that
\beq
\Rb_\so (s) = -\frac{1}{16\p^2} \; \Re\! \lb L(s) - L(m_\k^2) \rb 
\label{b.12}
\eeq

\begin{figure}[h] 
\hspace*{-25mm}
\includegraphics[width=0.35\columnwidth,angle=0]{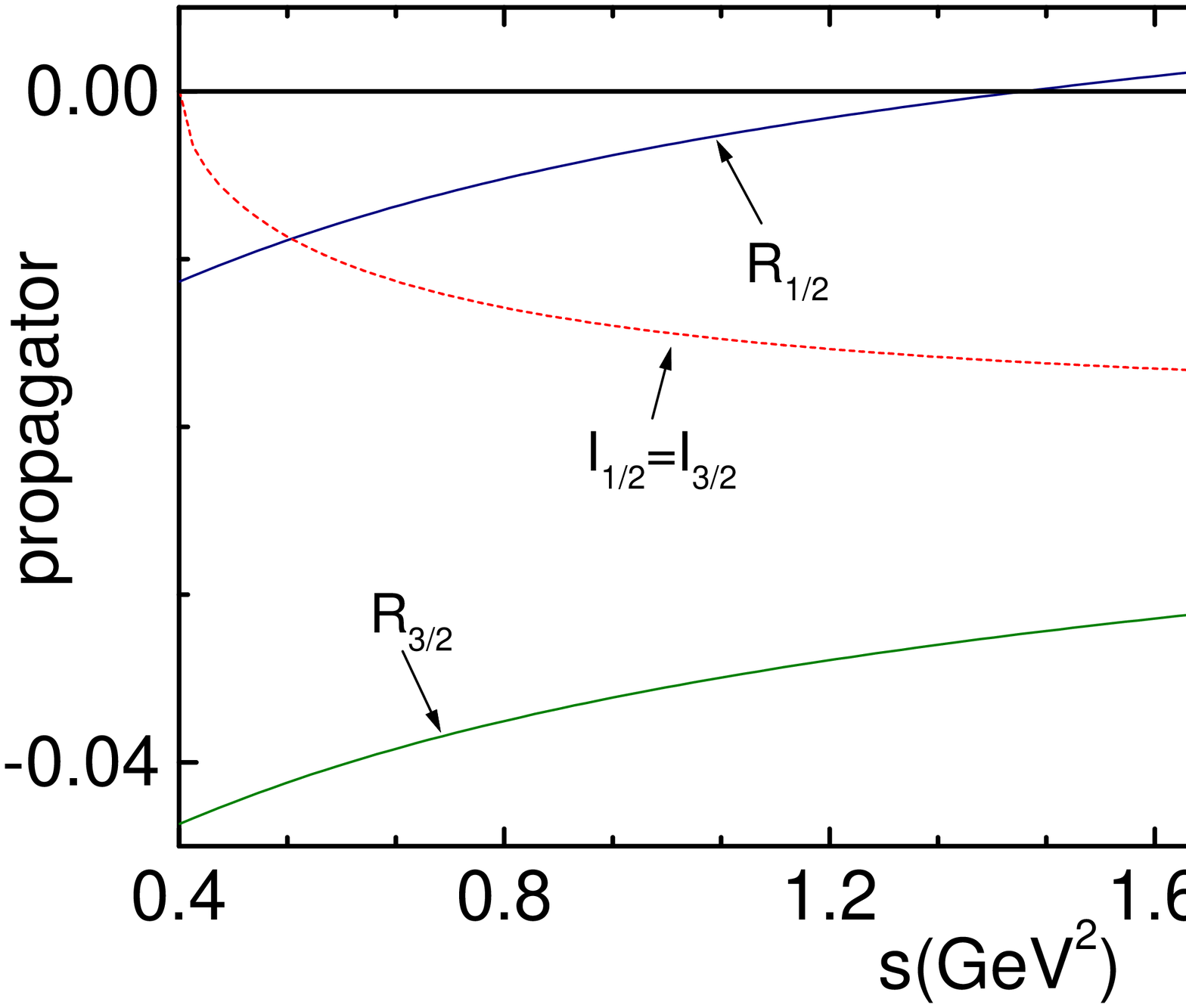}
\hspace*{25mm}
\includegraphics[width=0.35\columnwidth,angle=0]{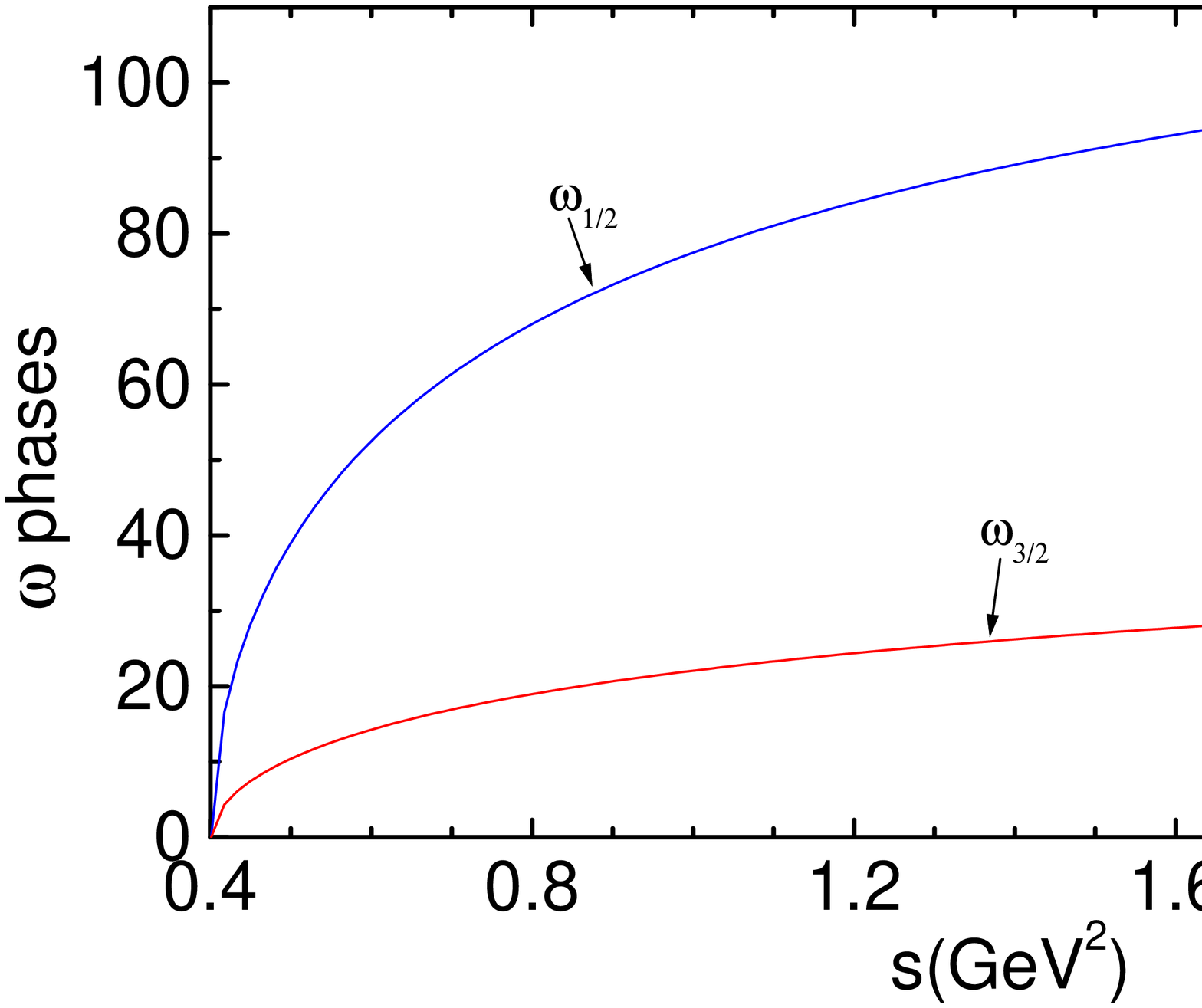}
\caption{(Color online) Left: real (continuous line) and imaginary 
(dashed line) components of the two-meson propagator;
right: propagator phases.}
\label{F20}
\end{figure}

\ni
and, by construction, $\Rb_\so(m_k^2)=0$.
In the $I=3/2$ channel, fit to data requires $c_\st\sim 140\;$GeV.
The functions $\Rb_I$, $I$ and $\o_I$ are shown in fig.\ref{F20}.

\section{background amplitudes}
\label{background}

The resonance-exchange amplitudes calculated in section \ref{kernel} have
the general form
\beq
T_x^a = - \, \frac{4}{F^4} \lc \frac{1}{x \sm m_a^2} \; 
[(c_d\,m_a^2 + C_\p^a) \; (c_d\,m_a^2 + C_K^a)]
+ c_d^2 \,(x \sp m_a^2) + c_d \, (C_\p^a \sp C_K^a) \rc \;,
\label{c.1}
\eeq

\ni
where $x = t, u$, and 
\bea
&& C_\p^\k = C_K^\k = 
- c_d\;(M_\p^2 \sp M_K^2) + c_m \;(4\,M_K^2 \sp 5\, M_\p^2)/6  \;,
\label{c.2}\\[2mm]
&& C_\p^0 = - 2\,(\ct_d \sm \ct_m) \, M_\p^2 \;,
\;\;\;\;\;\;\; C_K^0 = - 2\,(\ct_d \sm \ct_m) \, M_K^2 \;,
\label{c.3}\\[2mm]
&& C_\p^8 = - 2\,c_d \, M_\p^2  - c_m \,(2 M_K^2 \sm 11 M_\p^2)/6 \;,
\label{c.4}\\[2mm]
&& C_K^8 = 
- 2\, c_d\,M_K^2 + c_m \;(10 M_K^2 \sm M_\p^2)/6 \;.
\label{c.5}
\eea

Using the results for $S$-wave projection given in appendix \ref{kinematics},
the background amplitudes entering the kernels of section \ref{kernel} 
are written as 
\bea
&& B_{1/2} = 3\,B_s^\k/4 + B_t^0 - B_t^8/6 - B_u^\k/4 \;,
\label{c.6}\\[2mm]
&& B_{3/2} = B_t^0 - B_t^8/6 + B_u^\k/2 \;,
\label{c.7}\\[2mm]
&& \;\;\;\;\; B_s^\k = -\, \frac{4}{F^4} 
\lb c_d^2 \,(s \sp m_\k^2) + c_d \, (C_\p^\k \sp C_K^\k) \rb \;,
\label{c.8}\\[2mm]
&& \;\;\;\;\; B_x^a = - \, \frac{4}{F^4} \lc -\, \frac{1}{s\,\rho^2}\;
\ln \lb 1 \sp \frac{s\,\rho^2}{m_a^2\sm x_0 - s\,\rho^2/2} \rb \;
[(c_d\,m_a^2 + C_\p^a) \; (c_d\,m_a^2 + C_K^a)] \right.
\nn\\[2mm]
&& \left. \;\;\;\;\;\;\;\;\;
+ c_d^2 \,(x_0 \sp m_a^2) + c_d \, (C_\p^a \sp C_K^a) \rc \;.
\label{c.9}
\eea

\vspace{-5mm}

\begin{figure}[h]
\includegraphics[width=.45\columnwidth,angle=0]{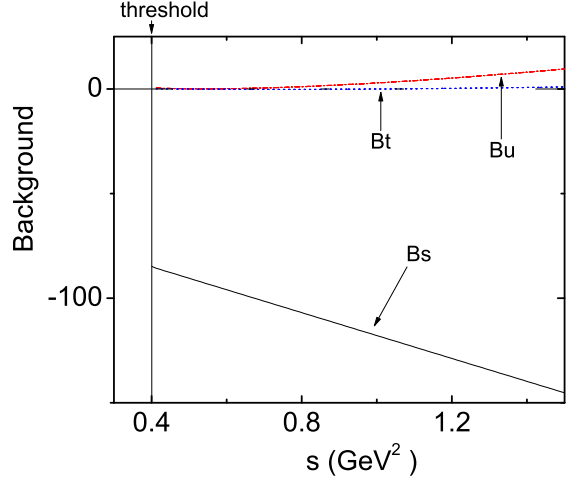}
\caption{Background amplitudes.}
\label{F21}
\end{figure}


\end{document}